\global\def\draftcontrol{0}

   \def\versionno{  } 

\catcode`\@=11 

\expandafter\ifx\csname draftcontrol\endcsname\relax\global\def\draftcontrol{0} 
\fi 

{\count255=\time\divide\count255 by 60 
\xdef\hourmin{\number\count255} 
\multiply\count255 by-60\advance\count255 by\time 
\xdef\hourmin{\hourmin:\ifnum\count255<10 0\fi\the\count255}} 
\def\draftdate{\number\month/\number\day/\number\year\ \ \ \hourmin } 

\newcommand\makepapertitle{\par

  \begingroup 
    \renewcommand\thefootnote{\@fnsymbol\c@footnote}%
    \def\@makefnmark{\rlap{\@textsuperscript{\normalfont\@thefnmark}}}%
    \long\def\@makefntext##1{\parindent 1em\noindent 
            \hb@xt@1.8em{%
                \hss\@textsuperscript{\normalfont\@thefnmark}}##1}%
     \newpage 
     \global\@topnum\z@   
     \@makepapertitle 
     \thispagestyle{empty}\@thanks 
  \endgroup 
  \setcounter{footnote}{0}%
  \global\let\thanks\relax 
  \global\let\makepapertitle\relax 
  \global\let\@makepapertitle\relax 
  \global\let\@thanks\@empty 
  \global\let\@author\@empty 
  \global\let\@date\@empty 
  \global\let\@title\@empty 
  \global\let\title\relax 
  \global\let\author\relax 
  \global\let\date\relax 
  \global\let\and\relax 
  \def\version{\let\version\@version\@gobble} 
} 
\def\@makepapertitle{%
  \newpage 
   \ifnum\draftcontrol=1 {} 
   \version\versionno 
   \vskip 5em%
   \else 
   \hfill\hbox to 3cm {\parbox{4cm}{\@pubnum}\hss}%
   \vskip 5em%
   \fi 
   \begin{center}%
   \let \footnote \thanks 
      {\hskip -0\textwidth \hbox to 1\textwidth%
        {\centerline{\Large\bf{\noindent\@title}}}}%
     \vskip 2em%
     {\normalsize
       \lineskip .5em%
       \begin{tabular}[t]{c}%
         \@author 
       \end{tabular}\par}%
     \vskip 1em%
     {\@bstract}%
     \end{center}%
     \vfill
     \@date%
     \vskip 1.5em%
   \par 
} 

\gdef\@pubnum{} 
\def\pubnum#1{%
  \gdef\@pubnum{#1}} 

\gdef\@bstract{} 
\def\Abstract#1{%
  \gdef\@bstract{%
   \parbox{\textwidth-0pc}{%
   \centerline{\bf Abstract}\penalty1000 
   \noindent
   \renewcommand\baselinestretch{1.0} 
   {#1}}} 
} 

\gdef\@email{}
\def\email#1{%
   \gdef\@email{%
   Email: {\tt #1}}
}

\def\ps@paper{\let\@mkboth\@gobbletwo%
     \ifnum\draftcontrol=1 
        \def\@oddfoot{\hbox to \textwidth{\tiny \versionno \hfil\tiny\draftdate}%
        \hskip -\textwidth \hbox to \textwidth{\hfil\rm\thepage\hfil}}%
     \else\def\@oddfoot{\hbox to \textwidth{\hfil\rm\thepage\hfil}} 
     \fi 
     \let\@evenfoot\@oddfoot 
} 

\def\body{\clearpage 
          \pagestyle{paper} 
        } 
\newenvironment{acknowledgments}{%
\vskip 3.25ex 
\noindent {\bf Acknowledgments} 
} 

\def\@version#1{\ifnum\draftcontrol=1 
\typeout{}\typeout{#1}\typeout{} 
\vskip3mm\centerline{\hbox{\fbox{\normalsize{\tt DRAFT -- #1 -- } 
                   {\draftdate}}}}\vskip3mm 
\fi} 
\let\version\@version 
\long\def\eqlabel#1{\ifnum\draftcontrol=1 
                    \tag@false  
                    \tag*{(\theequation) \hbox to -0.2cm{\hspace{0cm}\small{#1}\hss}} 
                    \refstepcounter{equation}  
                    \edef\@currentlabel{\theequation} 
                    \ltx@label{#1}          
                    \else 
                    \label{#1} 
                    \fi 
                    } 
\let\st@bibitem\@bibitem 
\let\st@lbibitem\@lbibitem 
\ifnum\draftcontrol=1 
  \def\@bibitem#1{%
    \st@bibitem{#1}\a@@label{#1}\ignorespaces} 
  \def\@lbibitem[#1]#2{%
    \st@lbibitem[#1]{#2}\a@@label{#2}\ignorespaces} 
  \def\a@@label#1{%
    \gdef\a@lab{\smash{\normalfont\small#1}} 
    \ifvmode 
      \if@inlabel 
        \global\setbox\@labels\hbox{%
          \llap{\a@lab\let\a@lab\relax 
                \kern\@totalleftmargin\kern\marginparsep}%
          \box\@labels}%
      \fi 
    \fi} 
\fi 

\documentclass[12pt,letterpaper]{article} 

\usepackage{amsmath,bm,amsfonts,amssymb,array,calc,amsthm,rotating,epsfig,psfrag} 
\usepackage[nosort]{cite} 
\usepackage{graphicx}

\renewcommand\baselinestretch{1.25} 
\renewcommand\section{\@startsection {section}{1}{\z@}%
                                   {-3.5ex \@plus -1ex \@minus -.2ex}%
                                   {2.3ex \@plus.2ex}%
                                   {\normalfont\large\bfseries}} 
\renewcommand\subsection{\@startsection{subsection}{2}{\z@}%
                                   {-3.25ex\@plus -1ex \@minus -.2ex}%
                                   {1.5ex \@plus .2ex}%
                                   {\normalfont\normalsize\bfseries}} 
\renewcommand\subsubsection{\@startsection{subsubsection}{3}{\z@}%
                                   {-3.25ex\@plus -1ex \@minus -.2ex}%
                                   {1.5ex \@plus .2ex}%
                                   {\normalfont\normalsize\it}} 
\renewcommand\paragraph{\@startsection{paragraph}{4}{\z@}%
                                   {-3.25ex\@plus -1ex \@minus -.2ex}%
                                   {1.5ex \@plus .2ex}%
                                   {\normalfont\normalsize\bf}} 
\renewcommand\subparagraph{\@startsection{subparagraph}{5}{\z@}%
                                   {-1.25ex\@plus -1ex \@minus -.2ex}%
                                   {0ex \@plus .2ex}%
                                   {\normalfont\normalsize\it}} 

\long\def\@makecaption#1#2{%
  \vskip\abovecaptionskip
  \sbox\@tempboxa{{\bf #1:} #2}%
  \ifdim \wd\@tempboxa >\hsize
    {\small\bf #1:} {\small #2}\par
  \else
    \global \@minipagefalse
    \hb@xt@\hsize{\hfil\box\@tempboxa\hfil}%
  \fi
  \vskip\belowcaptionskip}

\def\ie{{\it i.e.}} 
\def\eg{{\it e.g.}} 

\def\revise#1       {\raisebox{-0em}{\rule{3pt}{1em}}%
                     \marginpar{\raisebox{.5em}{\vrule width3pt\ 
                     \vrule width0pt height 0pt depth0.5em 
                     \hbox to 0cm{\hspace{0cm}{%
                     \parbox[t]{4em}{\raggedright\footnotesize{#1}}}\hss}}}}

\newcommand\nxt[1]  {\\\fnxt#1}

\def\calf         {{\cal F}}

\def\caln         {{\cal N}}

\def\calt         {{\cal T}} 
 
\def\calv         {{\cal V}}

\def\zet          {{\mathbb Z}} 

\def\del          {\partial} 
 
\def\ee           {{\it e}} 
\def\ii           {{\it i}} 
 
\def\tr           {\mathop{\rm Tr}} 
\def\Re           {{\rm Re\hskip0.1em}} 
\def\Im           {{\rm Im\hskip0.1em}}

\def\sqr#1#2{{\vcenter{\vbox{\hrule height.#2pt   
 \hbox{\vrule width.#2pt height#1pt \kern#1pt 
 \vrule width.#2pt}\hrule height.#2pt}}}}

\def\U{{\it U}}
\def\SU{{\it SU}}

\catcode`\@=12 

\begin{document} 


\title{Non-perturbative RR Potentials in the $\hat c=1$ Matrix Model}

\pubnum{%
NSF-KITP-03-100 \\ 
hep-th/0312021} 
\date{November 2003} 

\author{David J.\ Gross and Johannes Walcher \\[0.2cm] 
\it Kavli Institute for Theoretical Physics \\ 
\it University of California \\ 
\it Santa Barbara, CA 93106, USA \\[0.4cm] 
}


\Abstract{
We use the $\hat c=1$ matrix model to compute the potential energy $V(C)$ 
for (the zero mode of) the RR scalar in two-dimensional type 0B string 
theory. The potential is induced by turning on a background RR flux, 
which in the matrix model corresponds to unequal Fermi levels for the 
two types of fermions. Perturbatively, this leads to a linear runaway 
potential, but non-perturbative effects stabilize the potential, 
and we find the exact expression $V(C)=\frac{1}{2\pi}\int da\,\arccos 
\bigl[\cos(C)/\sqrt{1+\ee^{-2\pi a}}\bigr]$. We also compute the
finite-temperature partition function of the 0B theory in the 
presence of flux. The perturbative expansion is T-dual to the analogous
result in type 0A theory, but non-perturbative effects (which depend
on $C$) do not respect naive $R\to 1/R$ duality. The model can also 
be used to study scattering amplitudes in background RR fluxes.
}


\makepapertitle 

\body 

\version\versionno 


\section{Introduction}

The stabilization of moduli is an important problem on the way towards 
constructing realistic models in string/M-theory. It has been appreciated in 
recent years that a simple way of stabilizing moduli is to turn on fluxes in the 
compactification manifold. For example, in constructions based on the type 
IIB string or F-theory, turning on fluxes induces a superpotential which 
generically fixes all complex structure moduli \cite{gkp}. However, the K\"ahler
moduli and their associated RR partners remain untouched in this procedure,
and stabilizing them seems to require non-perturbative effects \cite{kklt}.
More precisely, all potentials that can be induced at the perturbative
level (by breaking supersymmetry) are runaway potentials, without any 
stable stationary point. But after including non-perturbative effects 
\cite{witten}, the potential can have (meta-)stable minima, of great 
cosmological interest \cite{kklt}.

In this paper, we study the potential for a particularly simple 
``modulus'', the zero mode of the Ramond-Ramond (RR) scalar $C$ of type 
0B ``non-critical'' string theory in two dimensions. This constant part 
of $C$ is related to a shift symmetry that is analogous to the gauge 
symmetries of $p$-form fields in higher-dimensional string theories. 
But while in higher dimensions, potentials for the periods of these 
$p$-form gauge fields are sometimes hard to compute due to the lack of 
a fully non-perturbative definition of string/M-theory, the advantage 
of the two-dimensional setup is that there exists an accessible 
non-perturbative definition of the theory, the so-called $\hat c=1$ 
matrix model \cite{tato,dkkmms}. We will therefore be able to compute 
everything exactly.

As we will see, the mechanism for generating the potential in the two-dimensional
0B context is essentially similar to the situation in higher dimensions. One can 
induce a potential for $C$ by turning on a RR background flux. Perturbatively, 
this potential would exhibit runaway, but this behavior is stabilized 
by non-perturbative effects. The exact expression is (see figure 
\ref{fig:pot} on page \pageref{fig:pot} for a graph of this function)
\begin{equation}
V(C) = \frac{1}{2\pi}\int\limits_{\mu-Q}^{\mu+Q} da 
\arccos\left[\frac{\cos C}{\sqrt{1+\ee^{-2\pi a}}}\right]\,,
\eqlabel{VofC}
\end{equation}
where $\mu$ is the tachyon background (worldsheet cosmological constant)
determining the string coupling, and $Q$ is the appropriately normalized
RR flux. The crucial step in understanding $V(C)$ is the identification, 
in the finite-$N$ matrix model, of the RR flux and the zero mode of $C$. 
Briefly, while $\mu$ corresponds to the distance of the Fermi level for the 
fermionic eigenvalues from the top of the inverted harmonic oscillator potential 
of the double-scaled matrix model, $Q$ corresponds to the difference of the Fermi 
levels for the even and odd modes of the inverted harmonic oscillator%
\footnote{This is the non-perturbative definition. Perturbatively, it can be
reduced to the difference of the Fermi levels on the two sides of the 
potential, or for left and right moving eigenvalues, depending on the sign
of $\mu$. This identification was proposed in \cite{dkkmms}.}. Moreover, $C$ 
is identified with the $\zet_2$-breaking boundary condition in the
asymptotic regions of the potential. We will give a careful derivation 
of these identifications in the upcoming section \ref{sec:defs}.

In section \ref{sec:pot}, we then turn to the computation of the potential
energy $V(C)$ as a function of the string coupling and background flux.
Since the finite piece \eqref{VofC} vanishes at $Q=0$, we will for 
completeness also compute the subleading contribution that vanishes
when we take the double-scaling limit.

We then give two other applications of our identifications of section
\ref{sec:defs}. In section \ref{sec:part}, we compute the finite-temperature
partition function for non-zero background flux $Q$, and compare with
the analogous results from the type 0A matrix model. We find that once the
0A flux is continued to imaginary values, the perturbative expansions of
the two results are T-dual to each other. The non-perturbative parts of 
the partition functions, however, are not mapped to each other under
naive $R\to 1/R$ duality. We speculate on the meaning of this result. 
On the other hand, our results are invariant under ``S-duality'', which
involves $\mu\to -\mu$ and dualization of $C\to\tilde C$. In addition,
we find a novel duality which exchanges NS and RR background 
$\mu\leftrightarrow Q$. In the matrix model, these dualities are simple 
particle/hole dualities for the even and odd modes of the potential.
Some of the relevant computations, such as the dependence of the
partition function on $C$, and the specialization of the results
to the ``self-dual'' radius $R=1$, can be found in the appendix.
In section \ref{sec:bos}, we study the bosonization of the matrix model 
eigenvalues when the Fermi levels for even and odd modes are not the 
same, and compute some S-matrix elements, following \cite{drsvw,kapustin}. 
We conclude in section \ref{sec:conc} with a list of open problems.

For reviews of matrix models useful for our present discussion, see 
\cite{gross,klebanov,gimo,polchinski}. More recent matrix model literature 
includes \cite{gv1,gv2,kms,gv3,gv4,gv5,gv6,gv7,gv8,gv9,gv10,gv11,gv12
,gv13,gv14,gv15,gv16}.

\section{The 0B matrix model and its deformations}
\label{sec:defs}

The matrix model which in the double-scaling limit describes two-dimensional
$\caln=1$ supergravity coupled to $\hat c=1$ matter with type 0B GSO
projection \cite{tato,dkkmms} is essentially similar to the conventional
or ``old-fashioned'' matrix model describing two-dimensional bosonic
gravity coupled to $c=1$ matter. The definition begins with the quantum 
mechanics of an $N\times N$ hermitian matrix $M$ with potential $V(M)$ which
has a quadratic maximum at $M=0$. In the singlet sector, this quantum mechanics 
can be reduced to the dynamics the matrix' eigenvalues $\lambda$ behaving as 
$N$ free fermions moving in the potential $V(\lambda)$. The continuum limit 
involves taking $N$ to infinity, sending Planck's constant to zero, and 
adjusting the various parameters such that the Fermi level approaches 
the top of the potential to order $\hbar$. The only parameters surviving the 
limit are the Fermi level $\mu$ and the curvature of the potential at the 
maximum. The bosonic and the supersymmetric model differ slightly 
but importantly in both.

In the ``old-fashioned'' way of thinking \cite{brka,dosh,grmi}, the 
expansion of the path-integral for finite $N$ captures the discretization 
of the two-dimensional string worldsheet. The double-scaling limit is 
taken in order to tune the model to criticality. In the ``reloaded'' 
form \cite{mcve}, one thinks of the quantum mechanics as arising 
holographically as the world-volume theory on a stack of a large 
number of unstable D0-branes. In this context, the curvature of the 
potential at the maximum is identified with the mass of the open string 
tachyon. It is $m^2=-1/\alpha'$ in the bosonic case and $m^2=-1/2\alpha'$ 
in the supersymmetric case.

Also, in the old days of the bosonic model, fermions were only filled on 
one side of the maximum, which can only make sense in the perturbative $\hbar$ 
expansion. For the supersymmetric version, we are now instructed to fill
both sides of the potential, and to interpret this non-perturbatively
stable model as the two-dimensional linear dilaton background of the
type 0B string. As in the bosonic case, collective excitations of the
fermionic eigenvalues are identified with perturbative string theory
degrees of freedom. In the bosonic case, there is only the center of mass 
of the string, which in two dimensions is a massless tachyon $T$. In the
supersymmetric 0B model, one has in addition a massless RR scalar $C$.
According to \cite{dkkmms,tato}, even fluctuations of the Fermi sea
correspond to $T$, and odd fluctuations correspond to $C$. In other words, 
the $\zet_2$ parity symmetry of the matrix model $M\to -M$ is identified 
with $(-1)^{F_L}$ in the string theory picture.

In this paper, we will be interested in studying aspects of the RR 
sector of this 0B matrix model. As we have mentioned in the introduction,
the RR scalar shares many interesting features with the RR $p$-form 
fields familiar from higher-dimensional string theories. For type 0 
theories in general, the spacetime effective action contains to lowest 
order terms of the form \cite{bega,klts,klts2,meor,dkkmms}
\begin{equation}
\frac{1}{8\pi}\int f(T)\; dC\wedge *dC
\eqlabel{kinetic}
\end{equation}
that couple the RR forms to the closed string tachyon $T$. Constraints
on the functions $f(T)$ can be obtained from the requirement of T-duality
\cite{meor}, and suggest taking $f(T)=\ee^{-2 T}$, as we will do, following
\cite{dkkmms}. In two dimensions, there is only one RR form, which is a 
(non-chiral) scalar boson. As usual, the zero mode of $C$ does not appear
in the action \eqref{kinetic}, and we have a one-parameter shift
symmetry. Similarly to higher dimensions, this perturbative gauge symmetry 
is violated non-perturbatively by instanton effects. Such instantons should 
generically generate a potential for $C$, and by analogy with the Yang-Mills 
$\theta$-parameter, one might expect that this potential has to lowest order 
the form $\ee^{-1/g_s} \cos C$. We will see that the details are in fact 
somewhat different in the present case.

The natural field strength associated with $C$ is $F=\ee^{-T} dC$,
satisfying the Bianchi identity and equation of motion
\begin{equation}
\begin{split}
d \bigl(\ee^{T} F\bigr) &=0 \,,\\
d \bigl(\ee^{-T} *\!F \bigr)&= 0 \,.
\end{split}
\eqlabel{Feom}
\end{equation}
Recall that to lowest order, the tachyon background is given by
\begin{equation}
T(\phi,t) = T(\phi) = \mu \ee^{\phi} \,,
\end{equation}
where $\phi$ is the spacelike Liouville direction supporting the linear
dilaton. In this background the linearized equations of motion 
\eqref{Feom} for $F$ admit the two linearly independent solutions 
\cite{dkkmms},
\begin{align}
F &= \ee^{-T} dt\,,
\eqlabel{electric} \\
F &= \ee^T d\phi\,,
\eqlabel{magnetic}
\end{align}
to which we shall refer to as the electric and magnetic solution,
respectively. Of course, since $C$ is the middle-dimensional form in the
two-dimensional 0B theory, we could equally well have chosen to 
describe this using the dual form $\tilde C$ related to $C$ by
$\ee^{-T} dC = F = *\tilde F = \ee^{T} *\!d\tilde C$, so that the
action for $\tilde C$ is (see \cite{meor} for some details on 
electric-magnetic dualities in type 0 theories)
\begin{equation}
\frac{1}{8\pi} \int \ee^{2T}\; d\tilde C\wedge *d\tilde C \,.
\eqlabel{kinmag}
\end{equation}
We also note that depending on the sign of $\mu$, one of the two solutions 
\eqref{electric}, \eqref{magnetic} decays at infinity $\phi\to\infty$, 
which as we recall is the asymptotic region where the matrix model lives.
For clarity, we will primarily discuss the case $\mu>0$, by which we mean 
that the Fermi level is below the top of the potential. In this case, 
only the electric solution \eqref{electric} is physically relevant.%
\footnote{Our conventions are slightly different from the ones used 
in \cite{dkkmms}, where the magnetic solution is physical when the Fermi 
level is below the top of the potential. In other words, the field $C$ 
of \cite{dkkmms} is our magnetic variable $\tilde C$.} But as is 
apparent from our discussion, one in facts expects that the theory 
enjoys an exact S-like duality $\mu\to -\mu$, $C\to\tilde C$. We will
confirm that after a proper identification of $C$ and $\tilde C$, all 
our results are invariant under this duality.

In \cite{dkkmms}, it was proposed to identify the generator of the shift 
symmetry of $C$ with the perturbatively (for $\mu>0$) well-defined difference 
in the number of fermions on the right and on the left of the maximum of 
the potential in the matrix model. One can justify this identification 
from the fact that instantons in spacetime are related to the tunneling 
of eigenvalues through the potential barrier in the matrix model. The 
instantons carry $C$-field charge and the tunneling changes the number of 
fermions on the left and right by one unit. Thus, a constant shift $C\to 
C+2\alpha$ shifts the phase of the wavefunction of fermions on the left 
of the potential by $-\alpha$ and the phase of the wavefunction of fermions 
on the right by $+\alpha$.

Another identification provided in \cite{dkkmms} is that the decaying 
solution \eqref{electric} is related to the difference of the Fermi level 
for fermions on the left and the right of the potential, which perturbatively  
is again a well-defined notion. But since we are now allowing eigenvalue  
tunneling, one is naturally led to wonder about the non-perturbative 
stability of this flux background. From the outset, we notice that 
the total flux associated with the solution \eqref{magnetic}, 
$\int_{-\infty}^\infty F_t\; d\phi\,$, as well as the energy, 
$\int F_t^2\;d\phi$, diverge as the volume of space. One the other 
hand, an instanton carries a quantized (electric) charge, and a single 
tunneling can only change the flux by a finite amount. Moreover, in the 
spacetime picture the instanton corresponds to a D-brane which is 
localized in the Liouville direction at infinite $\phi$, and has no 
zero mode. It is therefore not clear whether instantons are actually 
able to destroy the flux completely. We will show in this paper that 
while the instantons do affect the identification of the RR sector 
in the matrix model, the flux background \eqref{electric} is in fact 
perfectly stable. 

\begin{figure}[t]
\begin{center}
\epsfig{width=2.8in,height=1.5in,file=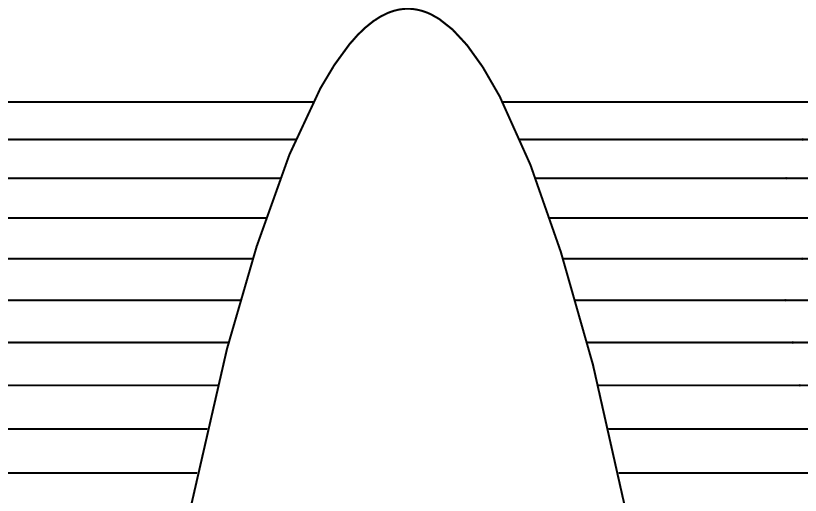}
\qquad
\epsfig{width=2.8in,height=1.5in,file=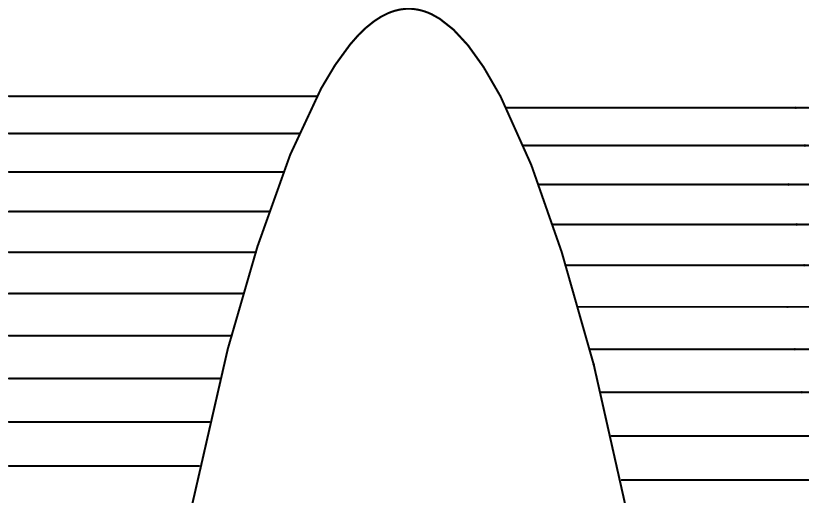}
\qquad
\end{center}
\caption{Perturbative energy levels in the inverted harmonic
oscillator potential (finite $N$). For symmetric boundary conditions 
($C=0$), the spectrum is doubly degenerate (left panel). Changing
$C$ shifts the phase of fermions on the left and right of the potential. 
This raises the levels on the left and lowers them on the right
(the right panel is for $C=1$, where the periodicity of $C$ is $2\pi$). 
The energy shifts cancel out in the sum and $C$ is perturbatively an exact 
symmetry in the double-scaling limit.}
\label{fig:pertel}
\end{figure}

We begin in the perturbative picture and consider the energy levels of
the matrix eigenvalues close to the top of the potential, see Fig.\
\ref{fig:pertel}. 
These energy levels are determined semiclassically by a systematic WKB 
expansion. To lowest order, we have the Bohr-Sommerfeld quantization 
condition for the $n$-th energy level $\epsilon_n$,
\begin{equation}
\pi n \hbar = \int^{\lambda_*} \sqrt{2(\epsilon_n-V(\lambda))} d\lambda \,,
\eqlabel{WKB}
\end{equation}
where we integrate up to the classical turning point $\lambda_*$
satisfying $V(\lambda_*)=\epsilon_n$. The boundary condition at infinity
depends on the details of the potential in this non-universal region,
but can at most contribute an overall phase to the WKB approximation.
If the boundary condition is $\zet_2$ symmetric, the energy levels 
on the left and right will match, $\epsilon_{L,n}=\epsilon_{R,n}\,$,
as shown on the left in Fig.\ \ref{fig:pertel}. We also record the 
familiar expression for the asymptotic density of states $\rho$, 
which is obtained from \eqref{WKB} upon approximating $V(\lambda)$ 
quadratically around $0$ with $V''(0)=-1$,
\begin{equation}
\rho=\Bigl|\frac{dn}{d\epsilon}\Bigr| = \frac{\ln 2\pi \hbar n}{2\pi\hbar} \,.
\eqlabel{asymrho}
\end{equation}
Shifting the phase of the wavefunction by $\alpha$ corresponds to adding
$\hbar \alpha$ on the RHS of \eqref{WKB}. In the convenient normalization 
$\alpha\sim C/2$, a shift of $C$ by $2\pi$ maps the $n$-th energy level
on the right to the $(n-1)$st and the $n$-th energy level on the left
to the $(n+1)$st. More precisely, we find from
\begin{equation}
\frac{dn}{dC} = \frac{1}{2\pi}  
\end{equation} 
that shifting $C$ changes $\epsilon_{L/R}$ by 
\begin{equation}
\frac{d\epsilon_{L/R}}{dC} = \pm \frac{1}{2\pi\rho}  \,.
\eqlabel{elshift}
\end{equation}
The two contributions cancel in the sum, so that if the energy levels are 
filled to the same Fermi level on the left and on the right, shifting
$C$ does not cost any energy, even before taking the double-scaling limit.
In the double-scaling limit, which involves $\rho$ diverging as $|\ln\hbar|$,
a finite difference of fermion number or a finite shift of $C$ does not
change the renormalized Fermi level $\mu$, which is measured from the 
top of the potential in units of $\hbar$. Moreover, since we neglect 
tunneling, the scattering amplitudes do not feel any mismatch of 
energy levels on the left and right of the potential. We conclude that $C$ 
is perturbatively an exact symmetry in the double-scaling limit.

\begin{figure}[t]
\begin{center}
\epsfig{width=2.8in,height=1.5in,file=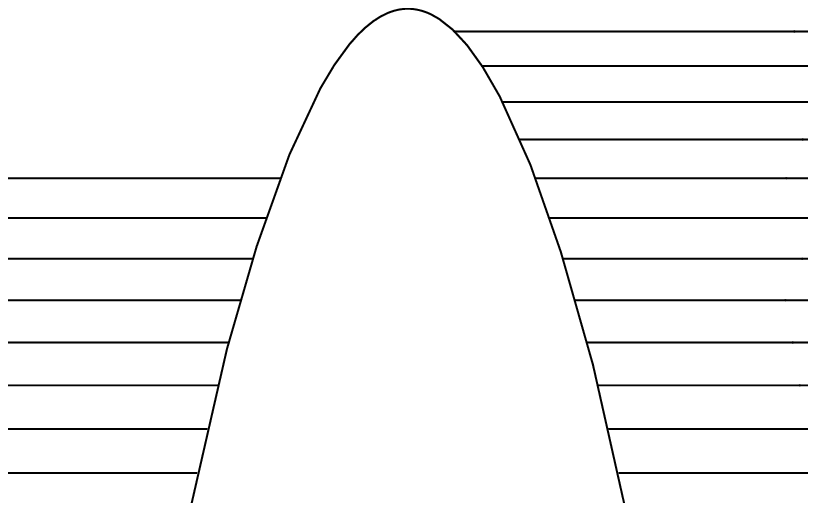}
\qquad
\epsfig{width=2.8in,height=1.5in,file=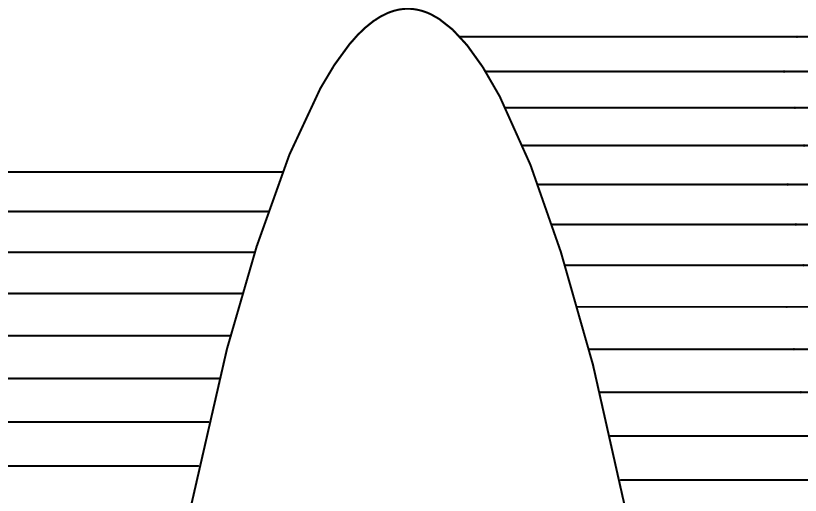}
\qquad
\end{center}
\caption{Perturbatively, we can fill the energy levels asymmetrically 
on the left and right of the potential, but keep the difference of
Fermi levels finite in the double-scaling limit. This induces a linear
potential for $C$.}
\label{fig:asym}
\end{figure}

These considerations suggest a simple way of inducing a non-vanishing
potential for $C$. The idea is to fill the eigenvalues asymmetrically, on 
the left up to $\mu_L$, and on the right up to $\mu_R$, and to keep the 
difference $\mu_L-\mu_R$ finite in the double-scaling limit, see
Fig.\ \ref{fig:asym}. We will parameterize $\mu_L=\mu+Q$ and $\mu_R=\mu-Q$.
In the double-scaling limit, non-zero $Q$ makes two contributions 
to the energy. There is first of all an infinite contribution that arises 
because we have lifted an infinite number (in the limit) of fermions 
by a finite amount. This contribution will diverge as usual as 
$|\ln\hbar|$, which is identified with the volume of space in the 
spacetime picture. But there is also a finite contribution, which 
depends on $C$. Indeed, upon integrating \eqref{elshift} we find the 
contribution
\begin{equation}
V(C) = \int\limits_{\mu_R}^{\mu_L} d\epsilon\;\rho\; \bigl(\epsilon_R(C)-
\epsilon_R(0)\bigr)
= -\frac{C Q}{\pi} \,.
\eqlabel{pertVofC}
\end{equation}
(The derivation is similar for the other sign of $Q$.) We note that this
term is non-zero but finite (not infinite) in the double-scaling limit.

One may wonder what kind of term in the spacetime action could 
capture a contribution of the form \eqref{pertVofC}. Following the
suggestion made in \cite{dkkmms}, we assume that $Q$ is related
to the flux of the RR one-form field strength, which in our conventions is
the electric solution \eqref{electric}, $F=\ee^{-T}dC \propto Q \ee^{-T}
dt$. We can then write \eqref{pertVofC} as 
\begin{equation}
V(C) \propto \bigl(CF_t(+\infty)-CF_t(-\infty)\bigr)
= \int d\phi\; \del_\phi\bigl(C F_t\bigr)\,,
\eqlabel{maga}
\end{equation}
which corresponds to a total derivative term in the action of the form
\begin{equation}
\int  d\bigl(CF\bigr) \,.
\eqlabel{magact}
\end{equation}
This term \eqref{magact} is reminiscent of Chern-Simons terms familiar from
higher-dimensional string theories. It would be interesting to see whether
one can check the presence of such a term from the spacetime or Liouville
point of view. The potential \eqref{VofC} we will compute can be viewed as 
the non-perturbative completion of the expression \eqref{magact}.

As we have mentioned, the potential energy \eqref{pertVofC} is linear in $C$
and does not have any stationary point. In addition, the flux background
in Fig.\ \ref{fig:asym} appears to be unstable to eigenvalue tunneling. We
will now give a qualitative description of how both $C$ and the flux
are stabilized once we include non-perturbative effects, and will present 
the quantitative calculations in the next section.

\begin{figure}[t]
\qquad
\psfrag{0}{\raisebox{.5cm}{$\;0$}}
\psfrag{pi}{\raisebox{.5cm}{$\pi$}}
\psfrag{mpi}{\raisebox{.5cm}{$-\pi$}}
\psfrag{tpi}{\raisebox{.5cm}{$2\pi$}}
\psfrag{mtpi}{\raisebox{.5cm}{$-2\pi$}}
\psfrag{C}{$C$}
\epsfig{width=5in,height=1.7in,file=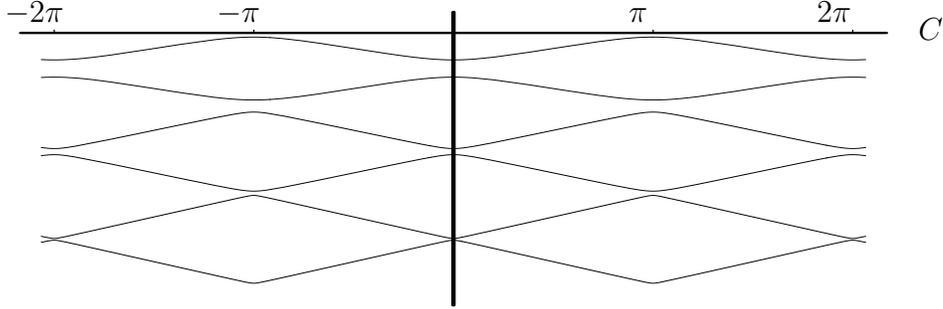}
\caption{Energy eigenvalues of the inverted harmonic oscillator
potential (at finite $N$) as a function of the boundary condition asymmetry
$C$. }
\label{fig:npev}
\end{figure}

The picture is quite simple (see Fig.\ \ref{fig:npev}). If we allow for tunneling, 
the energy levels on the left and right of the potential mix, and the 
true eigenstates of the inverted harmonic oscillator are linear combinations of 
the two. For symmetric boundary conditions ($C=0$), the eigenstates split into 
even and odd under $\lambda\to -\lambda$, with splitting (at finite $N$) of 
order $\ee^{-\pi a}$, where $a$ is the distance from the top of the potential 
in units of $\hbar$. Each eigenfunction is equally supported on the left and 
right of the potential. When $C$ is non-zero, the perturbative energy levels 
on the left and right do not match (see Fig.\ \ref{fig:pertel}) and mixing is 
suppressed. Therefore, the eigenfunctions on each side of the potential essentially
keep their identity, with small admixture from the other side. The eigenvalues
depend linearly on $C$, as in eq.\ \eqref{elshift}. When $C$ reaches $\pi$, 
the perturbative energy levels would cross, leading to eigenvalue repulsion. 
The eigenfunctions are again equally supported on both sides of the potential.
From $C=\pi$ to $C=2\pi$, the picture is the same with inverted sign, and the 
full periodicity is $C\to C+2\pi$. This dependence of the eigenvalues on 
$C$ is depicted in Fig.\ \ref{fig:npev}.

Even though for $C\neq 0$, the eigenstates of the inverted harmonic
oscillator are strictly speaking neither even nor odd, we will simply 
refer to the states by the parity they have once adiabatically 
continued to $C=0$. An equivalent notion is the parity of the
wavefunction under the combined reflection $\lambda\to -\lambda$,
$C\to -C$. But one has to keep in mind that except for small regions
of order $\ee^{-\pi a}$ around $C= 0$ and $\pi$, these wavefunctions 
are actually well supported on one side of the potential. 

Putting everything together, we are now led to the identification of
the electric RR flux background $Q$ with the difference of the Fermi 
levels for the even and odd eigenmodes of the inverted harmonic oscillator. 
Perturbatively, this definition reduces to the earlier one given in 
\cite{dkkmms}. To be specific, we will put the Fermi level for the 
even modes at $\mu_+=\mu+Q$ and for the odd modes at 
$\mu_-=\mu-Q$, and we assume for the time being that both $\mu_+$ and
$\mu_-$ are positive. The RR background thus corresponds to a natural 
deformation of the 0B matrix model, and is perfectly stable. Moreover, 
once $Q$ is non-zero, we induce a potential energy $V(C)$ for the 
zero mode of $C$, which is obtained by summing up the dependence of 
the unpaired even/odd eigenvalues from $\mu_-$ to $\mu_+$.
Before we turn to the computation of $V(C)$, we will give an 
independent heuristic argument for this identification of the RR 
flux in the matrix model.

According to \cite{mcve,dkkmms}, we can think of the matrix model 
holographically as the worldvolume theory on a stack of $N$ unstable 
D0-branes of type 0B string theory. In this context, the matrix $M$ 
can be identified with the open string tachyon $\calt$ of the 
D0-branes, and the inverted harmonic oscillator is the maximum of 
the tachyon potential $\calv(\calt)$. As is by now quite 
well established, open string tachyons on unstable D-branes also 
couple to RR fields via terms of the form \cite{tachyon}
\begin{equation}
\int \tr d\calf(\calt)\wedge C \,.
\eqlabel{CS}
\end{equation}
The purpose of this coupling is to give the correct RR charge to
a tachyonic kink interpolating between the two minima of $\calv$,
with the BPS condition $\Delta\calf=\int \calv$ relating charge
and mass. The couplings $\calv$ and $\calf$ are not known in general%
---in particular, they might depend on the closed string tachyon
in type 0B theory. However, all we need to retain presently is the 
behavior near the maximum of $\calv$. Then, since the constant
part of $\calf$ does not matter, all we need to know is that
while $\calv$ is generically an even function under $\calt\to 
-\calt$, $\calf$ is generically an odd function.

Eq.\ \eqref{CS} is also relevant for describing the behavior of 
unstable D-branes in an external field. Upon integration by parts, 
we obtain \cite{mcve}
\begin{equation}
\int \tr \calf(\calt) \wedge \hat F
\eqlabel{flux}
\end{equation}
where $\hat F=d C$ is the field strength associated with 
$C$, and we note again that $\calf$ is odd under $\calt\to
-\calt$.

All these facts strongly suggest that turning on RR flux should 
correspond holographically to a deformation of the matrix model 
which is odd under $M\to -M$ (or equivalently, $\lambda\to 
-\lambda$). In second quantized language, we are led to the
Hamiltonian
\begin{equation}
H=\int d\lambda\; 
\Bigl[ \frac12\frac{\del\Psi^\dagger(\lambda)}{\del\lambda}
\frac{\del\Psi(\lambda)}{\del\lambda}
-\frac{\lambda^2}{2}\Psi^\dagger(\lambda)\Psi(\lambda)
+\mu\Psi^\dagger(\lambda)\Psi(\lambda)
+ Q \Psi^\dagger(\lambda)\Psi(-\lambda)\Bigr] \,.
\eqlabel{second}
\end{equation}
The equations of motion that follow from \eqref{second} for the even
and odd parts of $\Psi$ immediately imply the identification
of even and odd Fermi levels $\mu_\pm=\mu\pm Q$ that we have claimed
above.

Finally, we would like to make a minor comment concerning the nature of 
the flux appearing in \eqref{CS} and \eqref{flux}. In general, type 0
string theories which are constructed with $\caln=2$ worldsheet 
supersymmetry contain twice as many RR fields as their type II cousins.
The middle-dimensional form has both electric and magnetic
degrees of freedom. This doubling arises because the RR ground
states of the NSR string with both even and odd worldsheet fermion
number contribute. In contrast, the present two-dimensional type 0B 
theory actually only enjoys $\caln=1$ worldsheet supersymmetry, and 
the corresponding type IIB model is dead (but see \cite{mmv}). (One is
tempted to call this the $\frac 02$B string.) It is no surprise, 
therefore, to find that there is only one kind of allowed RR 
background, which with our choice of variables is the electric 
background \eqref{electric}.

An analogous discussion holds for D-branes. Type 0B theories with 
$\caln=2$ worldsheet supersymmetry have two kinds of D-branes, 
conventionally called D$p+$ and D$p-$, each of which is charged
(electrically) under the appropriate combination of RR fields
\cite{bega,klts,klts2}. The middle-dimensional form has one 
electrically and one magnetically charged brane. In our case,
we have a reduction also of the number of branes because we only 
have $\caln=1$ worldsheet supersymmetry \cite{dkkmms}.

To complete the story, we therefore have to justify that the
RR form that couples to the tachyon on our unstable D0-branes
is indeed the electric variable, and not possibly the magnetic
one, $\tilde C$. Here, we can use the results of ref.\ 
\cite{thompson}, in which Sen's brane descent relations where 
studied for type 0 theories. According to \cite{thompson} there 
are two completely independent descent charts, one for the 
$+$ and one for the $-$ branes. If we assume that the electric 
flux is the difference of Fermi levels, the instanton which 
corresponds to tunneling of eigenvalues is charged electrically
under $C$. Therefore, since the instanton descends from the 
Euclidean kink on the unstable D0-brane, we conclude that the 
unstable D0-brane must indeed couple to the electric variable 
$C$, as we had needed for consistency.

\section{Computation of the potential}
\label{sec:pot}

In our notation, we will mostly follow the usual conventions of \cite{grkl,
moore,mpr}, except that we will call Planck's constant $\hbar$ instead of 
$1/\beta$, reserving $\beta$ for the inverse temperature appearing in 
section \ref{sec:part}. In the double-scaling limit, we take $\hbar\to 0$, 
keeping various other quantities fixed. The Schr\"odinger equation for the 
matrix model eigenvalue $\lambda$ is given by
\begin{equation}
\Bigl(-\frac{\hbar^2}{2}\del_\lambda^2 + V(\lambda) -\epsilon\Bigr)
\psi(\epsilon,\lambda)=0
\eqlabel{schrodinger}
\end{equation}
where $V(\lambda)$ is the matrix model potential. In the double-scaling 
limit, only the quadratic behavior near the maximum is relevant. 
We will make the convenient choice \cite{moore}
\begin{equation}
V(\lambda) = \frac 12(1-\lambda^2) \,,
\eqlabel{Vlambda}
\end{equation}
with infinite walls (Dirichlet boundary conditions on $\psi$) near
$\lambda=\pm 1$. For comparison with \cite{dkkmms}, we note that 
we have chosen units in which $\alpha'=1/2$.

It is convenient to introduce rescaled variables,
\begin{equation}
\lambda=\sqrt{\frac\hbar2} x\qquad\text{and}\qquad
\epsilon= \frac 12 - \hbar a \,,
\eqlabel{variable}
\end{equation}
in terms of which eq.\ \eqref{schrodinger} becomes
\begin{equation}
\Bigl(\del_x^2 + \frac{x^2}{4} - a\Bigr)\psi(a,x) = 0 \,.
\eqlabel{parabolic}
\end{equation}
This equation is solved by the parabolic cylinder functions called
$W(a,x)$, $W(a,-x)$ in the conventions of \cite{abst}. 

The zero mode of $C$ can easily be implemented by asymmetrically 
fine-tuning the positions of the walls in the limit $\hbar\to 0$. 
Explicitly, we impose Dirichlet boundary conditions $\psi(a,x_R)=
\psi(a,x_L)=0$, where
\begin{equation}
\begin{split}
x_L &= - \sqrt{\frac 2\hbar} + \sqrt{\frac\hbar 2}\, C \\
x_R &= \phantom{-} \sqrt{\frac 2\hbar} + \sqrt{\frac\hbar 2}\, C \,.
\end{split}
\eqlabel{def1}
\end{equation}
(This corresponds to $\lambda_L = -1+ \hbar C/2$ and 
$\lambda_R=1+\hbar C/2$.) One can easily check that shifting the walls
in this way shifts the perturbative energy levels as in 
\eqref{elshift}. The eigenvalue equation we have to solve then 
becomes
\begin{equation}
W(a,x_L) W(a,-x_R) = W(a,-x_L) W(a,x_R) \,.
\eqlabel{eveq}
\end{equation}
By utilizing the asymptotic expansion of $W(a,x)$ and $W(a,-x)$ 
given in \cite{abst}, we can reduce eq.\ \eqref{eveq} to the form
\begin{equation}
\frac 1k \sin\varphi_L \sin\varphi_R =
k \cos\varphi_L \cos \varphi_R \,,
\eqlabel{eveq2}
\end{equation}
up to terms of order $\hbar$. In \eqref{eveq2}, 
\begin{equation}
\begin{split}
\varphi_L &= \frac 14 x_L^2 - a\ln (-x_L) + \frac\pi 4 + \frac 12\Phi_2 \\
\varphi_R &= \frac 14 x_R^2 - a\ln x_R +\frac\pi 4 +\frac 12\Phi_2 \,,
\end{split}
\eqlabel{def2}
\end{equation}
and 
\begin{align}
k &= \sqrt{1+\ee^{2\pi a}} - \ee^{\pi a} \\
\Phi_2 &= \arg\Gamma\bigl(\frac12+ \ii a\bigr) \,.
\end{align}
A trigonometric identity brings \eqref{eveq2} into the form
\begin{equation}
\cos\bigl(\varphi_R-\varphi_L\bigr) = \sqrt{1+\ee^{-2\pi a}}\cos\bigl(
\varphi_R+\varphi_L\bigr) \,,
\eqlabel{eveq3} 
\end{equation}
which upon using the definitions \eqref{def2} and \eqref{def1} collapses
to
\begin{equation}
\cos C = \sqrt{1+\ee^{-2\pi a}} \cos 2\varphi_0 \,,
\eqlabel{eveq4}
\end{equation}
where 
\begin{equation}
2\varphi_0 = \varphi_L+\varphi_R = \frac12x_0^2 - 2a\ln x_0 +\frac\pi2+\Phi_2\,,
\eqlabel{def3}
\end{equation}
and $x_0=\sqrt{2/\hbar}$, and we need only retain terms up to that order in 
$\hbar$. Eq.\ \eqref{eveq4} determines the dependence of the eigenvalues $a$ 
of \eqref{parabolic} on $C$. Solutions of this equation come in pairs,
\begin{equation}
\varphi_0(a_{\pm}) = \mp\frac12\arccos\left[\frac{\cos C}{\sqrt{1+
\ee^{-2\pi a_{\pm}}}}\right] + n\pi \,,
\eqlabel{sol1}
\end{equation}
for each integer $n$. By explicitly constructing the wavefunctions associated
with these eigenvalues, one can check that the solutions $a_+/a_-$
correspond to even/odd modes, respectively, where we remind the reader
that we qualify modes as even or odd depending on their parity under the
combined reflection $(x,C)\to(-x,-C)$. We also recall that we can use
\eqref{sol1} to compute the density of states of the inverted harmonic
oscillator. Summing even and odd modes, the density of pairs is given 
up to terms of order $1/|\ln\hbar|$ in the double-scaling limit by
\begin{equation}
\rho(a) = \Bigl|\frac{dn}{da}\Bigr| = \frac1\pi \varphi_0'(a) \,.
\eqlabel{dos}
\end{equation}
In the limit, $\rho$ diverges as $|\ln\hbar|$.

To proceed, we parameterize the solutions of \eqref{sol1} as
\begin{equation}
a_\pm = a_0 + \frac 12\bigl(a_1 \pm a_2\bigr) \,,
\eqlabel{param}
\end{equation}
where we define $a_0$ by the property $\varphi_0(a_0)=n\pi$, \ie,
$a_0$ is independent of $C$, and $a_1$ and $a_2$ vanish in the
double-scaling limit as $1/|\ln\hbar|^2$ and $1/|\ln\hbar|$,
respectively, see eq.\ \eqref{result}. We can then expand \eqref{sol1} 
for fixed $a_0$ to find
\begin{equation} 
\varphi_0'\bigl(a_1\pm a_2\bigr) = \mp v - \frac12 v' a_2\,,
\eqlabel{expand}
\end{equation}
where everything is a function of $a_0$ and
\begin{equation}
v=v(a) = \arccos\left[\frac{\cos C}{\sqrt{1+\ee^{-2\pi a}}}\right] \,.
\eqlabel{vofa}
\end{equation}
Solving \eqref{expand} yields the leading corrections to the eigenvalues in 
the double-scaling limit,
\begin{equation}
\begin{split}
\varphi_0' a_2 &= - v(a_0) \\
\varphi_0' a_1 &= -\frac12 v' a_2 \,.
\end{split} 
\eqlabel{result}
\end{equation}
We have plotted this dependence of the eigenvalues \eqref{param} on $C$ 
in Fig.\ \ref{fig:npev} on page \pageref{fig:npev}. The plot shows $-a(C)$ 
for the first few eigenvalues closest to the top of the potential.

\begin{figure}[t]
\begin{center}
\psfrag{0}{$\;0$}
\psfrag{pi}{$\pi$}
\psfrag{mpi}{$-\pi$}
\psfrag{tpi}{$2\pi$}
\psfrag{mtpi}{$-2\pi$}
\psfrag{C}{$C$}
\psfrag{VC}{$V(C)$}
\epsfig{width=3in,height=1.5in,file=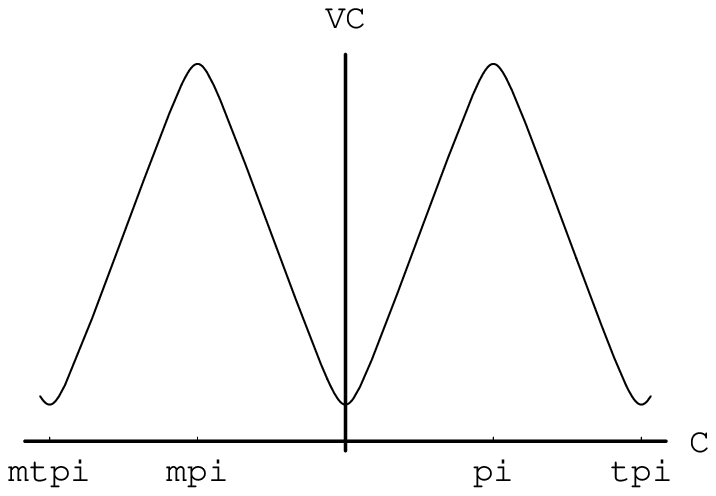}
\psfrag{VC}{}
\epsfig{width=3in,height=1.5in,file=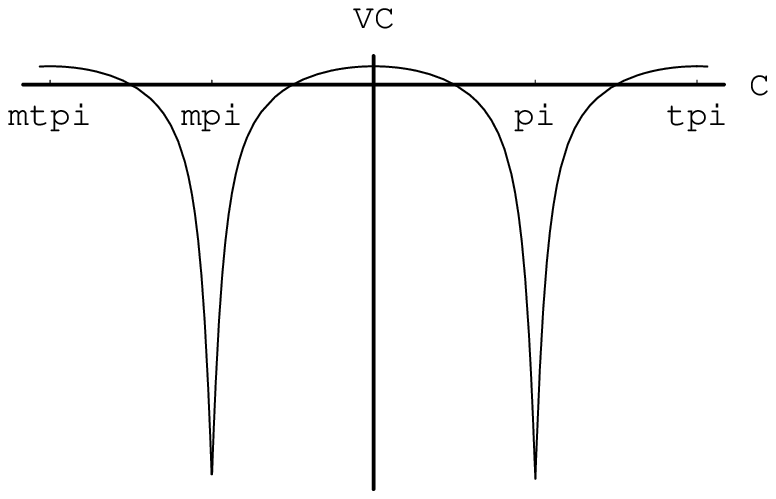}
\end{center}
\caption{The ground state energy of the matrix model depends on the
boundary condition asymmetry $C$. On the left, we plot the contribution
\eqref{final} which is finite in the double-scaling limit, for $\mu=.4$
and $Q=.1$. On the right, we show the subleading term \eqref{vanish} 
which goes to zero as $1/|\ln\hbar|$.}
\label{fig:pot}
\end{figure}

We are now in a position to evaluate the finite contribution to the ground
state energy of our matrix model that depends on the zero mode of the RR
scalar $C$. As we have explained in the previous section, we turn on RR 
flux by filling even modes up to $\mu_+=\mu+Q$ and odd modes up to 
$\mu_-=\mu-Q$ from the top of the potential, similarly to Fig.\
\ref{fig:asym} on page \pageref{fig:asym}. The ground state energy
is then given by summing over all the eigenvalues \eqref{sol1} in 
the parameterization \eqref{param} (recall that in the conventions of
this section, $a$ is counted from the top of the potential downwards)
\begin{equation}
\sum_{\mu_+}^\infty a_+ + \sum_{\mu_-}^\infty a_- =
\sum_{\mu}^\infty \bigl(2 a_0+a_1\bigr)
+\Bigl(\sum_{\mu_-}^\mu - \sum_{\mu}^{\mu_+}\Bigr) 
\bigl( a_0 +\frac12a_1\bigr)
- \sum_{\mu_-}^{\mu_+} \frac12a_2 \,.
\eqlabel{split} 
\end{equation}
The rationale behind this particular splitting is the following. The 
first term in \eqref{split} is independent of the flux. It has the 
usual divergences computed in \cite{grkl}. The second term depends 
on the flux, but is independent of $C$. It has a divergence which 
represents the bulk contribution of the background flux 
\eqref{electric}. We will analyze these contributions in more detail 
in the next section. The last term in \eqref{split} is the one 
of present interest. By plugging in the various definitions and using 
the density of states \eqref{dos}, we find that this term is finite in 
the double-scaling limit and given by
\begin{equation}
\begin{split}
V(C) &= -\int_{\mu_-}^{\mu_+} da_0\; \rho(a_0) \; \frac 12a_2(a_0) = 
-\frac{1}{2\pi} \int_{\mu_-}^{\mu_+}da\; \varphi_0'(a) a_2(a)
\\&=
\frac 1{2\pi} \int_{\mu-Q}^{\mu+Q}da\; 
\arccos\left[\frac{\cos C}{\sqrt{1+\ee^{-2\pi a}}}\right] \,,
\end{split}
\eqlabel{final}
\end{equation}
which is the main result of our paper. We show a plot of this
function in Fig.\ \ref{fig:pot}.

The term \eqref{final} is the only $C$-dependence that survives the
double-scaling limit. By curiosity, one may also ask for the next 
subleading term. We can easily extract this from the first term in 
\eqref{split} and find it to be
\begin{equation}
\frac{1}{4\pi|\ln\hbar|} \left. \biggl(\arccos\left[\frac{\cos C}
{\sqrt{1+\ee^{-2\pi a}}}\right]\biggr)^2\right|_{\mu}^\infty \,.
\eqlabel{vanish}
\end{equation}

In the perturbative limit $\mu\to\infty$, \eqref{final} obviously 
reduces to the result \eqref{pertVofC} found in the previous section. 
The only exceptions are the regions of order $\ee^{-\pi\mu}$ around 
the minimum and maximum of $V(C)$, where tunneling becomes dominant. 
It is curious to note that these non-perturbative effects actually 
lead to a large second derivative $V''\sim\ee^{\pi\mu}$ at
the minimum of the potential, in other words, the ``mass''
of the $C$-field appears to be of order $\ee^{1/g_s}$!
We strongly emphasize, however, that the quantity $V(C)$
we have computed cannot strictly be thought of as a mass 
term for $C$. In the spacetime interpretation, this potential energy 
is finite only after integrating over space, see eq.\ \eqref{maga}. 
Thus, the instanton contribution to the energy density vanishes, as 
could have been expected based on the fact that the instantons in 
question are D-branes localized at $\phi\to\infty$ in the Liouville 
picture \cite{ZZ,kms}. On the other hand, we note from the action
\eqref{kinetic} that the kinetic term for the $C$-field goes to 
zero at $\phi\to\infty$, so that even an infinitesimal potential
can conceivably lead to a dynamical stabilization of the constant
part of $C$. More precisely, if we assume that the potential
indeed comes from the instantons localized at infinite $\phi$,
we can imagine cutting out a finite part of the Liouville direction 
at large $\phi$ where the kinetic term for $C$ is very small.
In this region, there would be a finite energy density from $V(C)$,
which would stabilize the zero mode of $C$ (its value at the
cutoff) at the minimum of the potential.

Another interesting feature of $V(C)$ is that its minimum is 
at $C=0$ or $C=\pi$ depending on the sign of $Q$. This seems to
be a property of non-perturbative potentials for axion-like fields
also in higher dimensions, as for example the ones considered in
\cite{kklt}.

The term \eqref{vanish} is also interesting. First of all, its 
minimum is at $C=\pi$, which is somewhat unexpected. (For symmetry
reasons, the minimum can only be at $C=0$ or $C=\pi$.) Moreover, 
although it vanishes in the double-scaling limit, one can check 
that its second derivative with respect to $C$ is in fact infinite 
at the minimum $C=\pi$. Thus, even though $V(C)$ itself vanishes 
at $Q=0$, this subleading term seems to fix $C$ with high precision 
at $C=\pi$ also in the absence of flux. This is true in particular
if we imagine working with a cutoff Liouville direction.

We have here computed the exact answer for the potential from the 
matrix model point of view. It would naturally be extremely interesting 
to understand some of the features of the potential from the spacetime 
perspective.

Throughout the discussion, we have assumed that both Fermi levels
are well below the top of the potential, \ie, $\mu>0$, and $|Q|\ll 
\mu$). Only then is $C$ a perturbative symmetry of the matrix model.
When continuing $\mu$ to negative values (but keeping $|Q|\ll 
|\mu|$), $Q$ should be interpreted as the magnetic flux 
\eqref{magnetic}, or alternatively as electric flux of the dual
scalar $\tilde C$. Our formulas for $V(C)$ still make sense in 
that case, but the perturbative symmetry is the one associated 
with $\tilde C$. We compute the potential $V(\tilde C)$ in appendix 
\ref{app:S}.

\section{Finite temperature partition function and dualities}
\label{sec:part}

We will here follow the usual route developed in \cite{grkl} for
extracting finite thermodynamic quantities from the double-scaled
matrix model. In the matrix model (at finite $N$), we may consider 
the partition function
\begin{equation}
Z = \tr \ee^{-\beta (H-\mu N))} \,,
\eqlabel{partfun}
\end{equation}
where $\beta=1/T$ is the inverse temperature and $\mu$ is the chemical
potential. We note that we are here using rescaled variables that stay
fixed in the double-scaling limit. Thus, $H$ in \eqref{partfun} is the
Schr\"odinger operator \eqref{parabolic}, and $\mu>0$ is measured from
the top of the potential as in \eqref{variable}. We will parameterize
$\beta=2\pi R$, as is appropriate when thinking of \eqref{partfun} as
the path-integral with Euclidean time direction compactified on a 
circle of radius $R$. 

An important quantity for defining the thermodynamic limit of interest 
for string theory is the mean particle number
\begin{equation}
\Delta = \langle N\rangle=\frac1{Z}\tr N \ee^{-\beta(H-\mu N)}
=\frac{\del F}{\del\mu}\,,
\end{equation}
where $F= \ln Z/\beta$ is the grand-canonical free energy. More 
precisely, the double-scaling limit which defines non-critical string 
theory involves $\hbar\to 0$, $\Delta\to\infty$, keeping the
chemical potential $\mu$ fixed. Since we are dealing with free
fermions, it is easy to write $\Delta$ in the thermodynamic limit as
\begin{equation}
\Delta = \int_{-\infty}^\infty da\; \rho(a) f(a) \,,
\eqlabel{Delta}
\end{equation} 
where $\rho$ is the density of states and $f$ is the usual Fermi
distribution function
\begin{equation}
f(a)=f(a,\mu) = \frac{1}{1+\ee^{2\pi R(\mu-a)}} \,.
\end{equation}

The diverging expression \eqref{Delta} for $\Delta$ can be studied 
more conveniently by taking two derivatives with respect to $\mu$. 
Using that $\del f/\del \mu = -\del f/\del a$ and $f'(a)\to 0$ for 
$a\to\pm\infty$, one finds
\begin{equation}
\frac{\del^2\Delta}{\del \mu^2} = \int da\; \rho'(a) f'(a) \,.
\eqlabel{dtDelta}
\end{equation}

We briefly review the case of the bosonic string. Using \eqref{dos}
and the integral representation of the digamma function ($\digamma$ 
is often known as $\psi$, but we have already used that letter),
\begin{equation}
\rho'(a) = \frac{1}{2\pi}\Im \digamma'\bigl(\frac12+\ii a\bigr)
= \frac{1}{2\pi} \int_0^\infty dt\;\frac{t/2}{\sinh t/2}
\sin a t \,,
\eqlabel{rhob}
\end{equation}
as well as
\begin{equation}
f'(a) = \frac{\pi R}{2}\frac{1}{\cosh^2\pi R(\mu-a)} \,,
\end{equation}
we can write \eqref{dtDelta} as a ``Fourier integral''
\begin{equation}
\frac{\del^2\Delta}{\del \mu^2} =
\frac{1}{2\pi\mu} \int_0^\infty  dt\; 
\frac{t/2R\mu}{\sinh t/2R\mu} \;
\frac{t/2\mu}{\sinh t/2\mu} \,
\sin t\,.
\eqlabel{bosonic}
\end{equation}
For comparison with \cite{dkkmms}, we note that we are here 
working with units in which $\alpha'=1$, as follows from the fact
that the curvature of the matrix model potential \eqref{Vlambda} 
is $V''=-1$, which is identified with the mass of the tachyon 
$m^2=-1/\alpha'$. The result \eqref{bosonic}
is invariant under T-duality,
\begin{equation}
R\to R'=\frac{\alpha'}{R}\,, \qquad\qquad \mu\to \mu'=\frac{R}
{\sqrt{\alpha'}}\mu\,.
\eqlabel{trafo}
\end{equation}
More precisely, recalling that $\Delta$ is related to the partition
function via three derivatives with respect to $\mu$ and a factor of $R$, 
we see that
\begin{equation}
Z(R,\mu) = Z(R',\mu')
\end{equation}
holds for the bosonic matrix model.

When studying the 0B theory, we may introduce two different
chemical potentials $\mu_+$ and $\mu_-$ for even and odd modes, 
respectively. As before, $\mu_\pm=\mu\pm Q$, where $Q$ is related 
to the RR flux background as we have explained in section 
\ref{sec:defs}. For the computation of the partition function, 
one has to be slightly careful to split the various contributions 
in the right way. Starting at finite $N$, we have
\begin{equation}
\Delta = \sum \bigl( f(a_+,\mu_+) + f(a_-,\mu_-)\bigr)\,,
\eqlabel{naive}
\end{equation}
where the sum is over all eigenvalues which are given by the
solutions of \eqref{eveq4}. One way to evaluate \eqref{naive}
is to work with different ``densities of states'' for even and
odd modes, defined by differentiating \eqref{sol1}. On the other 
hand, there is an even mode for every odd mode, so that the two
densities should actually be the same. The answer is identical, 
but we find it cleaner to proceed by using the parameterization 
in \eqref{param}. We obtain
\begin{equation}
\begin{split}
\Delta &= \sum\bigl( f(a_0,\mu_+) + f(a_0,\mu_-)\bigr)
+\sum \frac12 a_2 \bigl(f'(a_0,\mu_+) - f'(a_0,\mu_-)\bigr) \\
&=\int da\; \rho(a) \bigl(f(a,\mu_+) +f(a,\mu_-)\bigr)
+ \int da\; \frac1{2\pi} v(a) \bigl(f'(a,\mu_+)-f'(a,\mu_-)\bigr) \,, 
\end{split}
\eqlabel{eee}
\end{equation}
where we use the same density of states $\rho$ as in the bosonic 
case \eqref{dos}. We can now again apply two derivatives to 
\eqref{eee}, and use the representation, 
\begin{equation}
\frac{1}{2\pi} v''(a) = 
\frac{\pi}{4}\frac{\sinh\pi a}{\cosh^2\pi a} = \frac{1}{2\pi}
\int_0^\infty dt\; \frac{t/2}{\cosh t/2} \sin a t \,.
\eqlabel{rep}
\end{equation}
In this expression, we have set $C$ to the minimum of its potential
($0$/$\pi$ depending on the sign of $Q$). It is possible to
compute the Fourier transform of $v'(a)$ also for general $C$,
but we relegate the somewhat cumbersome expressions to the
appendix. We now find
\begin{equation}
\begin{split}
\frac{\del^2\Delta}{\del\mu^2} &= 
\frac 1{2\pi} \int_0^\infty \frac{t/2R}{\sinh t/2R}\left[ 
\frac{t/2}{\sinh t/2} \Im \bigl(\ee^{\ii\mu_+ t} + 
\ee^{\ii\mu_- t}\bigr) +
\frac{t/2}{\cosh t/2} \Im \bigl(\ee^{\ii\mu_+ t} -
\ee^{\ii\mu_- t}\bigr)
\right] \\
&=\frac{1}{\pi} \int_0^\infty dt\;
\frac{t/2R}{\sinh t/2R}\;\left[
\frac{t/2}{\sinh t/2} \sin \mu t\,\cos Qt +
\frac{t/2}{\cosh t/2} \cos \mu t\,\sin Qt
\right] 
\,.
\end{split}
\eqlabel{obpart}
\end{equation}
For $Q=0$, and remembering that $\alpha'=1/2$ in our conventions for 
the supersymmetric model, this reduces to the expression given 
in \cite{dkkmms}. The reader might wonder about the second term in 
the square brackets in \eqref{obpart}, which appears to be odd
under $Q\to -Q$, whereas the partition function should naively
not depend on the sign of the flux. The resolution of this puzzle
is that we have not included the dependence on the zero mode of $C$,
the minimum of whose potential switches from $C=0$ to $C=\pi$
as we change the sign of $Q$. Taking this into account makes
\eqref{obpart} symmetric under $Q\to -Q$. (See eq.\ \eqref{obpartfull}
in the appendix.)

The result \eqref{obpart} is the complete non-perturbative result for 
the partition function as a function of $\mu$, $Q$ (and $C$). One can 
think of the first term in the square brackets as the non-perturbative 
resummation of the asymptotic series in $1/\mu^2$ that is defined 
in string perturbation theory. In fact, this asymptotic series
is essentially the sum of the two series 
\begin{equation}
\Delta''_{\rm pert} (\mu,Q) = \Delta''_{\rm pert}(\mu_+)
+\Delta''_{\rm pert}(\mu_-) \,,
\end{equation}
where the perturbative expansion of $\Delta_{\rm pert}''(\mu)$ is 
identical to the one of the bosonic string \cite{grkl}
\begin{equation}
\begin{split}
\Delta_{\rm pert}''(\mu) &=
\frac 1{2\pi} \int_0^\infty dt\;\frac{t/2R}{\sinh t/2R}\;
\frac{t/2}{\sinh t/2} \;\Im \ee^{\ii\mu t} \\
&\sim \frac {-1}{2\pi} \sum_{m=0}^\infty \frac1{\mu^{2m+1} (4R)^{m}}
(2m)!\sum_{k=0}^m |2^{2k}\!-\!2|\, |2^{2m\!-\!2k}\!-\!2|
\frac{|B_{2k}|\, |B_{2m\!-\!2k}|}{(2k)!\, (2m\!-\!2k)!} R^{m-2k} \,,
\end{split}
\end{equation}
where the $B_{2k}$ are the Bernoulli numbers. On the other hand,
one can see that the second term in square brackets in \eqref{obpart}
is analytic in $\mu$ and its asymptotic expansion in $1/\mu^2$ 
vanishes, thus indicating its non-perturbative nature. To make
this explicit, we write
\begin{equation}
\Delta''_{\rm np}(\mu,Q) = \Delta''_{\rm np}(\mu_+) - 
\Delta''_{\rm np}(\mu_-)
\end{equation}
with
\begin{equation}
\Delta''_{\rm np}(\mu) = \frac 1{2\pi} \int_0^\infty dt\; 
\frac{t/2R}{\sinh t/2R}\;
\frac{t/2}{\cosh t/2} \;\Im \ee^{\ii\mu t} \,.
\eqlabel{poles}
\end{equation}
The essential feature of this integral is that the integrand is even 
under $t\to -t$. We can then close the integration contour in the 
upper/lower half plane (depending on the sign of $\mu$) and write the 
integral as a sum over poles. Explicitly we have for $\mu>0$,
\begin{equation}
\Delta''_{\rm np}(\mu) = \sum_{n=1}^\infty \left\{
(-1)^{n+1}\; \frac{n^2\pi^2 R^2}{\cos n\pi R} \;\ee^{-2\pi n\mu R} +
(-1)^{n+1}\; \frac{(n-{\textstyle\frac12})^2\pi^2/R}
{\sin(n-{\textstyle\frac12})\pi/R} \;\ee^{-2\pi (n-{\textstyle\frac12})\mu}
\right\} \,,
\eqlabel{instantons}
\end{equation}
and a similar expression for $\mu<0$. We see from \eqref{instantons}
that there are two types of non-perturbative contributions. One
type comes from the poles of the $\frac1{\cosh t/2}$ factor in \eqref{poles}
and begins at order $\ee^{-\pi\mu}$.\footnote{Interestingly, for 
$C\neq 0$, these contributions start at order $\ee^{-2\pi\mu}$,
see appendix \ref{app:V}.} The spacetime interpretation
of these contributions are D-instantons, \ie, ZZ-brane \cite{ZZ} in the 
Liouville direction together with Dirichlet boundary condition in the 
Euclidean time direction. The other type of contributions comes from the 
poles of the thermal factor in \eqref{poles} and begins at order 
$\ee^{-2\pi R\mu}$. Obviously, the spacetime interpretation of this 
term is a D0-brane winding around the compact Euclidean time direction,
\ie, ZZ-brane together with Neumann boundary condition on the
circle. It is interesting to note that something special happens
to the structure of the non-perturbative series when the radius
$R$ is rational. In that situation, the two types of poles collide
and combine into a single contribution (this limit is built into 
\eqref{instantons} automatically). It would be nice to understand 
this structure from the Liouville point of view. The relation 
between ZZ-branes and non-perturbative contributions in the {\it 
bosonic} matrix model has also been discussed in \cite{gv6}.

We stress that it is really sensible to write an instanton expansion
for this second term in \eqref{obpart} because it is odd under $Q\to 
-Q$ and hence vanishes at $Q=0$. The corresponding amplitudes vanish 
in perturbation theory, and receive contributions only from the 
instantons. This property is in distinction to the first term
in \eqref{obpart}, whose perturbative expansion is non-zero.
From this asymptotic expansion, one can merely reconstruct that 
its non-perturbative ambiguities set in at order $\ee^{-2\pi\mu R}$ 
and $\ee^{-2\pi\mu}$, but it is not sensible to write an instanton 
expansion for these ambiguities.

With these results in hand, we now proceed to study various dualities
that we expect our theory to enjoy. We begin with T-duality. The
partition function of type 0A theory has been computed in \cite{dkkmms}
using the matrix model. It is found to be
\begin{equation}
\frac{\del^2\Delta}{\del\mu^2} =
\frac 1{2\pi} \int_0^\infty dt\; \frac{t}{\sinh t}\;
\frac{t/2R}{\sinh t/2R} \sin \mu t\, \ee^{- |q|t} \,,
\eqlabel{oapart}
\end{equation}
where $q$ is the integer flux parameter of the 0A theory measuring 
the net number of D0-branes. Is is easy to see that the results for ß
$q=Q=0$ are related by the simple T-duality
\begin{equation}
Z_A(R_A,\mu_A,q=0) = Z_B(R_B,\mu_B,Q=0)\,,
\end{equation}
where the parameters are related by
\begin{equation}
R_A = \frac{1}{2R_B} = \frac{\alpha'}{R_B}\,, \qquad 
\mu_A = 2R_B\mu_B = \frac{R_B}{\sqrt{\alpha'/2}}\, \mu_B \,.
\eqlabel{weird}
\end{equation}
It is worthwhile to point out that \eqref{weird} is a slightly
non-standard transformation rule on the string coupling, which in the
usual type II context transforms as in \eqref{trafo}. It would 
be interesting to understand why the string coupling has to be
normalized in a different way in this two-dimensional type 0 context.

If we want to check T-duality also for non-zero flux, the form of the
above expressions suggests that we should try to Wick rotate $q\to 
\ii Q$ and supplement \eqref{weird} with
\begin{equation}
\ii q = Q_A = Q_B\frac{\mu_A}{\mu_B}= \frac{\sqrt{2\alpha'}}{R_A} \,.
\eqlabel{fluxtrafo}
\end{equation}
The fact that the flux transforms with a factor of $\ii$ is quite 
suggestive of a timelike T-duality. Namely, recall that the flux-parameter 
$Q$ in type 0B is for an electric flux, which is a timelike one-form. 
Rotating to Euclidean time, the flux becomes imaginary. T-duality turns
the one-form into a ``zero-form flux'', which is really just the dual 
of the 2-form flux of 0A. But now undoing the analytical continuation
does not give back the factor of $\ii$, so that the flux stays imaginary. 
We will then take \eqref{weird} and \eqref{fluxtrafo} as a hypothesis for
T-duality, and compute the partition function for the 0A model at 
imaginary  $q=\ii Q$. We will be able to match the perturbative
expansions of the two theories, but we will not be able to find 
agreement at the non-perturbative level.

In the type 0A matrix model introduced in \cite{dkkmms}, the 
Schr\"odinger equation for the eigenvalues (the analogue of 
\eqref{parabolic}) is (see also \cite{jeyo,dkr} for early studies
of this matrix model potential)
\begin{equation}
\Bigl(\frac1x\del_xx\del_x +\frac{x^2}{4}-\frac{q^2}{x^2}-a\Bigr)\psi(a,x) = 0\,.
\eqlabel{problem}
\end{equation}
Writing $z=\frac{\ii x^2}{2}$ and $\psi=z^{q/2} \ee^{-z/2} f$, we obtain 
the confluent hypergeometric equation
\begin{equation}
\Bigl(z\del_z^2 +(1+q-z)\del_z - \bigl(\frac 12+\frac q2-
\frac{\ii a}{2}\bigr)\Bigr) f =0 \,,
\end{equation}
with well-known solutions. The problem with the analytical continuation 
$q\to \ii Q$ comes from the boundary condition at $x=0$. Indeed, the 
two solutions of \eqref{problem} behave as $\psi\sim x^\alpha$ with 
$\alpha^2=q^2$ near $x=0$. We see that for $q^2\ge 1$, only one of the
solutions is normalizable, $\int_0 x^{2\alpha+1}<\infty$ and there is 
no need for a boundary condition. For $q^2<1$, the singularity is 
of limit-circle type and we need an extra boundary condition to make
sure that the differential operator \eqref{problem} is self-adjoint.
The usual procedure for $q^2=0$ is to require that the Laplace operator
in the plane be self-adjoint. But for $q^2=-Q^2<0$, the solutions
are actually oscillatory near $x=0$. The possible real boundary 
conditions come in a one-parameter family,
\begin{equation}
\psi\sim \ee^{\ii \delta/2} x^{\ii Q} + \ee^{-\ii \delta/2} x^{-\ii Q}
\qquad \text{as $x\sim 0$, with $\delta$ real.}
\eqlabel{boundary}
\end{equation}
Using the traditional Kummer notation for the confluent hypergeometric
function, we can write the solution of \eqref{problem} satisfying 
the boundary conditions \eqref{boundary} as
\begin{equation}
\psi(a,x) = 
\textstyle
\ee^{\ii \delta/2} x^{\ii Q} \ee^{-z/2} 
\Phi\bigl(\frac12+\frac{\ii Q}2-\frac{\ii a}2, 1+\ii Q;z\bigr)+
\ee^{-\ii \delta/2} x^{-\ii Q} \ee^{-z/2} 
\Phi\bigl(\frac12-\frac{\ii Q}2-\frac{\ii a}2, 1-\ii Q;z\bigr)\,.
\end{equation}
By working through the asymptotics of the $\Phi$ function, one
finds that this solution behaves near $x\to\infty$ as
\begin{equation}
\psi(a,x)\sim {\it const.}\;\frac{1}{x}
\cos\bigl(\frac{x^2}4- a \ln \frac{x}{\sqrt{2}}-\frac\pi4-\varphi\bigr) \,,
\end{equation}
where the phase shift $\varphi$ is given by
\begin{equation}
\varphi = \arg\left[ 
\ee^{\ii (\delta+ Q\ln 2)/2}\frac{\Gamma(1+\ii Q)}
{\Gamma\bigl(\frac12+\frac{\ii Q}2 + \frac{\ii a}2\bigr)}
\ee^{-\pi Q/4} +
\ee^{-\ii (\delta+Q\ln 2)/2}\frac{\Gamma(1-\ii Q)}
{\Gamma\bigl(\frac12-\frac{\ii Q}2 +\frac{\ii a}2\bigr)}
\ee^{\pi Q/4}
\right]
\end{equation}
After absorbing an $a$-independent phase into $\delta$, this reduces to
\begin{equation}
\varphi =\arg\left[ 
\frac{\ee^{\ii \delta/2}\ee^{-\pi Q/4}}
{\Gamma\bigl(\frac12+\frac{\ii Q}2 + \frac{\ii a}2\bigr)}
+
\frac{\ee^{-\ii \delta/2}\ee^{\pi Q/4}}
{\Gamma\bigl(\frac12-\frac{\ii Q}2 +\frac{\ii a}2\bigr)}
\right]\,.
\eqlabel{shift}
\end{equation}
It seems hard to find an explicit integral representation for the
density of states $\frac{1}{\pi} \frac{d\varphi}{da}$ similar to
\eqref{rhob}. But noticing that
\begin{equation}
\left|\frac{\Gamma\bigl(\frac12-\frac{\ii Q}2 +\frac{\ii a}2\bigr)}
{\Gamma\bigl(\frac12+\frac{\ii Q}2 +\frac{\ii a}2\bigr)}
\ee^{\ii\delta}\ee^{-\pi Q/2}\right|^2 =
\frac{\ee^{\pi a} + \ee^{-\pi Q}}{\ee^{\pi a}+\ee^{\pi Q}}\,,
\eqlabel{good}
\end{equation}
we can expand the problem for large $\mu\gg 1$. We write $\varphi=
\varphi_1+\varphi_2$ as the sum of two terms
\begin{equation}
\begin{split}
\varphi_1 &= \textstyle
-\frac12\left[\arg\Gamma\bigl(\frac 12+\frac{\ii Q}2+
\frac{\ii a}2\bigr) + 
\arg\Gamma\bigl(\frac12-\frac{\ii Q}2+\frac{\ii a}2\bigr)\right] \\
\varphi_2 &= \arg[w + w^{-1}]\,,
\end{split}
\eqlabel{two}
\end{equation}
where $w= \bigl[\ee^{\ii\delta} \ee^{-\pi Q/2} 
\Gamma\bigl(\frac12-\frac{\ii Q}2+\frac{\ii a}2\bigr)/
\Gamma\bigl(\frac12+\frac{\ii Q}2+\frac{\ii a}2\bigr)
\bigr]^{1/2}$ has absolute value exponentially close to one.
The first term in \eqref{two} gives after the usual steps
\begin{equation}
\frac{\del^2\Delta}{\del\mu^2}\approx \frac{1}{2\pi}\int_0^\infty
dt\; \frac{t}{\sinh t}\;\frac{t/2 R}{\sinh t/2R} \sin\mu t\cos Qt\,,
\end{equation}
which is precisely the T-dual of the first term in \eqref{obpart},
as we have claimed. This fact that the asymptotic expansions 
of the two partition functions agree to all order in perturbation 
theory can also be seen by analytically continuing directly 
the result \eqref{oapart}. Indeed, as we have seen above, 
only the even part of $\ee^{-q t}$ will lead to a perturbative
series. Thus, the perturbative expansion can be obtained by
replacing the exponential with $\cosh qt$, which leads to a
convergent integral for sufficiently small $q$. Now rotating
$q\to\ii Q$ directly shows the agreement with the first term 
in \eqref{obpart}. But of course, it is safer to first analytically
continue the problem as we have done above.

What about the non-perturbative contributions? On the 0A side, these
come from the second term in \eqref{two}. We find for large $a\gg Q$
\begin{equation}
\varphi_2 \sim \ee^{-\pi a} \sinh \pi Q \, \tan 
\textstyle \bigl(\frac\delta 2- \frac Q2 \ln a\bigr) \,,
\eqlabel{dependence}
\end{equation}
where we have again absorbed all $a$-independent phases into 
$\delta$. We see that the first non-perturbative contribution to
the 0A partition function at imaginary flux $Q=\ii q$ is of order
$\sim \ee^{-\pi \mu} \sinh\pi Q$ and depends in a somewhat strange 
way on the boundary condition $\delta$. This result can indeed be
matched with the order of the first correction in \eqref{instantons} 
after using the T-duality transformation rules \eqref{weird}.
The fact that we need the additional boundary condition at $x=0$ 
is suggestive of trying to match it with $C$. We have computed
the corresponding dependence of the first instanton 
contribution on $C$ from the formulas in the appendix but 
have not been able to match this with the dependence of 
\eqref{dependence} on $\delta$. And the prefactors of the 
exponentials do not show an exact match either.

Moreover, one has to ask for the presence of the second type
of non-perturbative contributions, which begin at order 
$\ee^{-\pi \mu}$ on the 0B side. On the 0A side, this would
be a contribution of order $\ee^{-\pi\mu R}$ and has to come
from the thermal density factor $1/(1+\ee^{2\pi R(\mu-a)})$.
But these contributions start at order $\ee^{-2\pi R\mu}$,
in disagreement with expectations. It therefore seems unlikely 
to us that the naive prescriptions that we have followed above 
can reveal an exact T-duality between 0A and 0B, once the flux 
is turned on and non-perturbative effects are accounted for. 

Non-perturbative violations of T-duality have been noticed 
before, as for example in \cite{aspl}. In this paper, the 
duality group of the heterotic string on K3$\times T^2$ 
is studied. It is argued that the failure of naive T-duality 
at the non-perturbative level should be viewed not as a breakdown 
of perturbative intuition, but rather as a deformation of the 
latter. In the present context, this would suggest to deform
the T-duality relations \eqref{weird} by terms of order 
$\ee^{-\pi \mu}$ and $\ee^{-\pi \mu R}$ such that the partition
functions agree exactly. It would be worthwhile to study this
proposal further.

Before leaving the subject, we also wish to clarify what we mean by
``the type 0A matrix model at imaginary $q$''. The definition
of the 0A matrix model in \cite{dkkmms} uses rectangular 
$(N+q)\times N$ dimensional complex matrices, and a 
$U(N+q)\times U(N)$ gauge group which removes the phase of the
``eigenvalues''. The holographic intuition behind this definition
is that the matrix model is the worldvolume theory of a large
number of $D0$ and $\overline{D0}$ branes, and $q$ is the net 
$D0$-brane charge. Leaving aside the fact that this theory is
rather unphysical to study at imaginary $q$, it is not even
clear how to define the matrix model mathematically! 
What we have in mind here is an alternative definition of
the 0A matrix model, in which the $q$ leftover $D0$-branes
have been dissolved into background flux. Similarly to the
discussion at the end of section \ref{sec:defs}, the tachyon
$\calt$ on a general $D_p\overline{D}_p$ system, which is a 
complex field, couples to the RR $C^{(p-1)}$ form potential 
such that a vortex of $\calt$ carries one unit of lower-dimensional
charge. This corresponds to a term $\oint d \arg \calt \wedge 
C^{(p-1)}=\int d (d \arg \calt)\wedge C^{(p-1)}$ in the
worldvolume action. Integrating by parts, this becomes
$\int d\arg\calt\wedge F^{(p)}$. Thus, in the type 0A theory in 
two dimensions, if we turn on two-form flux $F^{(2)}=\ast
F^{(0)}\propto q$, we obtain a term $q \int\frac{d}{dt}
(\arg\lambda)$ in the quantum mechanics of the complex 
eigenvalues $\lambda$. It is easy to see that after gauge fixing, 
such a term corresponds precisely to a shift of the angular 
momentum in the eigenvalue plane by $q$ units, just as in 
\eqref{problem}. This way of defining the 0A model does not pose 
any obvious obstacles to $q\to \ii Q$, except for the fact that 
it might be difficult to define the model at finite $N$ because 
the Hamiltonian is unbounded from below.

Another interesting duality \cite{dkkmms} of the type 0B theory 
in two dimensions is ``S-duality'' $\mu\to -\mu$ and $C\to\tilde 
C$, as we have mentioned several times already. Our results
for the partition function at $C=0$ are obviously
invariant under $\mu\to -\mu$, $Q\to -Q$ which is simply
particle-hole duality of the matrix model eigenvalues. It
is somewhat less clear what becomes of our non-perturbative 
results involving the zero mode of $C$. While the formulas
do not suffer any obvious pathology, $C$ as described in 
section \ref{sec:defs} is not a perturbative symmetry for fermions 
above the top of the potential. For such states, the perturbative 
energy levels are left/right moving, and non-perturbative effects 
are reflections off the potential. The perturbative symmetry 
associated with these energy levels is precisely the zero mode 
of the dual variable $\tilde C$, and one can check that the 
non-perturbative dependence of eigenvalues on $\tilde C$ is 
precisely as in \eqref{sol1}, with $a\to -a$. This is discussed 
in detail in appendix \ref{app:S}.

We can also notice yet another duality from our matrix model
computations. The partition function (which is the third
integral of \eqref{obpart} with respect to $\mu$) is also
invariant under $\mu_-\to -\mu_-$, \ie, applying particle-hole
duality to the odd modes only. In terms of $\mu$ and $Q$, 
this corresponds to exchanging NS and RR backgrounds 
$\mu\leftrightarrow Q$, a rather intriguing duality indeed! 
The T-dual of this type of duality has been noticed before 
in the 0A context in \cite{kapustin}. It is much more manifest 
in the 0B model. 

To summarize, the 0B matrix model depends on two parameters 
$\mu$ and $Q$, or equivalently $\mu_+$ and $\mu_-$, but models
which differ only by a sign of $\mu_\pm$ are equivalent to
each other. If $|Q|\ll |\mu|$, the matrix model is the 
non-perturbative definition of a 0B non-critical string 
theory. Depending on the sign of $\mu$, $Q$ is interpreted
as either electric or magnetic flux with corresponding
perturbative gauge symmetry $C$ or $\tilde C$. It would 
be interesting to understand whether there is also a 
perturbative string theory description in the case $|\mu|
\ll |Q|$. An analog of the perturbative gauge symmetry does 
not seem to exist in that case.

\section{Bosonization and amplitudes in RR flux background}
\label{sec:bos}

According to our identification in section \ref{sec:defs}, the matrix
model with different Fermi levels for even and odd modes provides a
dual description of type 0B string theory in the RR flux background
\eqref{electric}. This formulation is extremely simple and makes it 
possible to compute scattering amplitudes and a complete S-matrix for
this string theory. In this section, we will describe the formalism 
for studying these amplitudes, leaving a detailed study of their 
properties for future work. Our approach follows the older works 
\cite{moore,mpr} on the bosonic string and the more recent paper
\cite{drsvw} on the type 0B model (with vanishing RR flux). The use 
of matrix models to study non-critical strings in flux backgrounds
was first proposed in the context of type 0A in \cite{kapustin}.

In relating matrix quantum mechanics to spacetime physics in two 
dimensions, one identifies collective excitations of the Fermi sea of 
matrix eigenvalues with the bosonic fields of the corresponding string 
theory. This identification includes a non-local field redefinition
which is implemented by the so-called ``leg-pole factors'' in the
scattering amplitudes, but we can ignore this subtlety for our
purposes. For the 0B model, it was proposed in \cite{tato,dkkmms}
to identify the even fluctuations of the Fermi surface with
the tachyon field $T$ and odd fluctuations with the RR scalar
$C$. In \cite{drsvw}, it was pointed out that unitarity of
the S-matrix requires the inclusion of ``solitonic sectors''
into the Hilbert space of asymptotic states. These solitonic
sectors are created by two additional bosonic fields, which we will
call $S^+$ and $S^-$, which carry $C$-field charge, and have no
perturbative spacetime interpretation.

More precisely, as in \cite{moore}, one begins by introducing 
second quantized fermion fields
\begin{equation}
\Psi(t,\lambda) = \int_{-\infty}^\infty \ee^{\ii\omega t}
b_\epsilon(\omega) \psi^\epsilon(\omega,\lambda) \,,
\eqlabel{field}
\end{equation}
where the $\psi^\epsilon(\omega,\lambda)$ are a complete basis of 
solutions of the classical field equation \eqref{schrodinger} or 
\eqref{parabolic}, and where we identify $\omega=-a$ to conform
with other conventions in the literature. For fixed $\omega$,
different such bases are appropriate in different situations. Of
primary interest for defining scattering amplitudes are in/out 
bases which are defined by requiring specific asymptotic behavior.
We will denote solutions which are in/outgoing on the left and 
right of the potential by subscripts L and R, 
respectively. The 'in' and 'out' bases are related, as usual,
by the S-matrix, 
\def\bl{b_{\rm L}}
\def\br{b_{\rm R}}
\begin{equation}
\begin{pmatrix} \bl^{\rm out} \\ \br^{\rm out}\end{pmatrix}
=\begin{pmatrix} R & T\\ T & R \end{pmatrix}
\begin{pmatrix} \bl^{\rm in} \\ \br^{\rm in} \end{pmatrix} \,,
\eqlabel{Smatrix}
\end{equation}
where we have chosen the phases of L and R modes such that the 
$\zet_2$ symmetry of the potential is manifest (\ie, $S$ commutes 
with $\left(\begin{smallmatrix} 0 & 1\\1&0\end{smallmatrix}
\right)$). By textbook methods, one finds the reflection and 
transmission coefficients to be
\begin{equation}
R(\omega) = \frac{\ii}{\sqrt{2\pi}} \ee^{-\pi \omega/2} \Gamma\bigl(\frac12-
\ii \omega\bigr)\,, \qquad
T(\omega) = \ii \ee^{\pi \omega}R(\omega)  \,.
\end{equation}
We stress that \eqref{field} is the full ``interacting field'', where 
in the present case the only interactions are with the external 
potential, and does not depend on the conventional choice of basis. 
For example, for the purpose of defining the vacuum, it will be mandatory 
to introduce other bases of modes with fixed parity.

But let us for the moment focus on the asymptotic left/right bases.
Both in and out modes yield the usual algebra of creation and
annihilation operators,
\begin{equation}
\begin{split}
\{\bl(\omega),\bl^\dagger(\omega')\} &= \delta(\omega-\omega')\\
\{\br(\omega),\br^\dagger(\omega')\} &= \delta(\omega-\omega')\\
\text{and} \qquad \{b_\epsilon^\#(\omega),b_\epsilon^\#(\omega')\} &=0
\qquad\qquad \text{for all other combinations.}
\end{split}
\eqlabel{fermionic}
\end{equation}
To complete the definition of the theory, we have to fix a representation
of this algebra, as we will do a little later. But independent of this
representation, we can introduce bosonic operators as fermion
bilinears such as
\begin{equation}
\begin{split}
a_{\rm LL}(\omega) &= \int_{-\infty}^\infty \bl^\dagger(\xi)
\bl(\xi+\omega) \\
a_{\rm RR}(\omega) &=  \int_{-\infty}^\infty \br^\dagger(\xi)
\br(\xi+\omega)  \,.
\end{split}
\eqlabel{llrr}
\end{equation}
As usual, these bosonic operators satisfy the relations of a current 
algebra\footnote{Actually, obtaining this algebra requires 
appropriately regulating in the UV, as is the case in our situation.}
\begin{equation}
\begin{split}
[a_{\rm LL}(\omega_1),a_{\rm LL}(\omega_2)] &=
\omega_1 \delta(\omega_1+\omega_2) \\
[a_{\rm RR}(\omega_1),a_{\rm RR}(\omega_2)] &=
\omega_1 \delta(\omega_1+\omega_2)  \,.
\end{split}
\eqlabel{current}
\end{equation}

In \cite{tato,dkkmms}, it was proposed to identify the spacetime tachyon
field $T$ with the even combination $a_{\rm LL}+a_{\rm RR}$ and
the RR scalar with the odd combination $a_{\rm RR}-a_{\rm LL}$.
But as pointed out in \cite{drsvw}, the bosonic S-matrix computed
with these fields cannot possibly be unitary. Indeed, it is easy to
see that we can define two further fermion bilinears,
\begin{equation}
\begin{split}
a_{\rm LR}(\omega) &= \int_{-\infty}^\infty \bl^\dagger(\xi)
\br(\xi+\omega) \\
a_{\rm RL}(\omega) &=  \int_{-\infty}^\infty \br^\dagger(\xi)
\bl(\xi+\omega)  \,,
\end{split}
\eqlabel{lrlr}
\end{equation}
which complete \eqref{current} to a $\U(2)$ current algebra
\begin{equation}
\begin{split}
[a_{\rm RR}(\omega_1),a_{\rm RL}(\omega_2)] &=
a_{\rm RL} (\omega_1+\omega_2) \\
[a_{\rm LL}(\omega_1),a_{\rm RL}(\omega_2)] &=
-a_{\rm RL} (\omega_1+\omega_2) \\
[a_{\rm RL}(\omega_1),a_{\rm RL}(\omega_2)] &=
0 \\
[a_{\rm RL}(\omega_1),a_{\rm LR}(\omega_2)] &=
a_{\rm RR} (\omega_1+\omega_2) - a_{\rm LL}(\omega_1+\omega_2)
+ \omega_1 \delta(\omega_1+\omega_2)\,,  \qquad{\it etc.}
\end{split}
\eqlabel{utwo}
\end{equation}
In this algebra, the tachyon $T\sim a_{\rm LL}+a_{\rm RR}$ is one
of the $\U(1)$'s and the RR scalar $C\sim a_{\rm RR}-a_{\rm LL}$ 
together with the ``soliton operators'' $S^+\sim a_{\rm LR}$ and 
$S^-\sim a_{\rm RL}$ generate an $\SU(2)$ algebra. The main point
of \cite{drsvw} is that because the interactions (scattering of 
fermions off the potential) do not respect this symmetry algebra, 
it is inconsistent to restrict the Hilbert space of asymptotic states 
to a fixed charge under the $\U(1)^2$ currents \eqref{current}.

Going back to the fermionic picture, we recall that we still have to
specify the vacuum of our theory. This amounts to splitting the 
fermionic algebra \eqref{fermionic} into creation and annihilation 
operators, in other words, the choice of a Fermi level. As we have 
explained in detail in section \ref{sec:defs}, the only non-perturbatively 
stable possibility is to choose separately Fermi levels for the even and odd
modes, related to the L/R modes by
\begin{equation}
b_{\pm}(\omega) = \frac 1{\sqrt{2}}\bigl(\bl(\omega)\pm\br(\omega)\bigr) \,.
\eqlabel{lrpm}
\end{equation} 
Thus, we choose $\mu_+=\mu+Q$, $\mu_-=\mu-Q$, and assume for
simplicity that $\mu_+<0$ and $\mu_-<0$. The vacuum $|0\rangle$ is 
defined by\footnote{In the conventions of this section, $\mu<0$
when the Fermi levels are below the top of the potential.}
\begin{equation}
\begin{split}
b_\pm(\omega)|0\rangle &= 0 \qquad \omega>\mu_\pm \,,\\
b_\pm^\dagger(\omega)|0\rangle &=0 \qquad \omega<\mu_\pm \,.
\end{split}
\eqlabel{vacuum}
\end{equation}
As we have mentioned, for $\mu_+=\mu_-$, the bosons defined in 
\eqref{llrr} have the interpretation of collective excitations of 
the Fermi surfaces which are coming in from the left or from the 
right, while the off-diagonal bosons \eqref{lrlr} create solitonic
sectors. For $Q\neq 0$, we loose a good semiclassical picture of 
the ground state, so this interpretation is not so obvious anymore.

Finally, we write out explicitly the two-point functions computed
using the definitions \eqref{llrr}, \eqref{lrlr} with respect
to the vacuum defined in \eqref{vacuum}. The formulae are simplest
in terms of $a_{\pm\pm} = b_\pm^\dagger b_\pm$, which can be related
to $T$, $C$, $S^+$ and $S^-$ by working through the definitions
\eqref{lrpm}. We find
\begin{equation}
\begin{split}
\langle a_{++}^{\rm out} (\omega_1) a_{++}^{\rm in}(\omega_2)\rangle
&= \delta(\omega_1+\omega_2) \int_{\mu_+-\omega_1}^{\mu_+}
d\xi_1\; R_+(\xi_1+\omega_1)R_+^*(\xi_1) \\
\langle a_{+-}^{\rm out} (\omega_1) a_{-+}^{\rm in}(\omega_2)\rangle
&= \delta(\omega_1+\omega_2) \int_{\mu_--\omega_1}^{\mu_+}
d\xi_1\; R_-(\xi_1+\omega_1)R_+^*(\xi_1) \\
\langle a_{+-}^{\rm out} (\omega_1) a_{+-}^{\rm in}(\omega_2)\rangle
&= 0\,, \qquad {\it etc.} \,,
\end{split}
\eqlabel{twopoint}
\end{equation}
where the $R_\pm$ are the eigenvalues of the free fermion S-matrix
\eqref{Smatrix}
\begin{equation}
R_\pm = R\pm T = \frac{1}{\sqrt{2\pi}} \Gamma\bigl(\frac12-\ii\omega\bigr)
\Bigl(\ii\ee^{-\pi\omega/2} \mp \ee^{\pi\omega/2}\Bigr)
\end{equation}

\section{Open Problems}
\label{sec:conc}

In this paper, we have studied aspects of the RR sector of type 0B 
string theory in two dimensions, using the matrix model as a
non-perturbative definition. The crucial step was the precise
identification, following \cite{dkkmms}, of the RR flux background
as the difference of Fermi levels for even and odd modes of the
inverted harmonic oscillator potential of the double-scaled
matrix model. We have also identified the zero mode of the RR
scalar $C$ in the finite-$N$ matrix model as the mismatch of the
perturbative (in $\hbar$) energy levels on the two sides of the 
potential.

We have then given three applications of this identification of RR
fields in the matrix model. Firstly, we have computed the 
potential energy for the zero mode of $C$, which is induced by 
turning on flux, and stabilized by non-perturbative effects.
Our result is the complete answer to this question, the analogue
of which in higher dimensions is of great interest for the 
construction of realistic string theory vacua. Secondly, we
have computed the finite-temperature partition function of the
0B string in the flux background. We have seen that T-duality
with the 0A model is realized at the perturbative level, but
appears to be violated by non-perturbative effects. Thirdly, 
we have outlined how one can use the matrix model to study 
scattering amplitudes of 0B strings in RR backgrounds, following 
\cite{moore,mpr,kapustin,drsvw}. Understanding strings in RR 
backgrounds is also an important problem in higher-dimensional 
string theories.

To conclude this paper, we list here a few open questions which
have been raised by our results. 
\nxt
Quite obviously, it would be extremely interesting to understand 
the spacetime or string theory computation of our result for 
the potential $V(C)$. There, the origin of the potential are 
non-perturbative effects associated with D-instantons, and it
would be interesting to see how they contribute to $V(C)$.
\nxt
More generally, it would be interesting to compute some other
terms in the effective action and to see the structure of 
non-perturbative effects, which could in particular provide 
guidance for similar questions in higher dimensions.
\nxt
It would be nice to understand better what happens to T-duality
with 0A for non-vanishing flux.
\nxt
The matrix model gives complete control over interpolations 
between different dual perturbative string theories. It would
be worthwhile to understand what happens at the transition 
region, \eg, when $\mu_-$ goes through zero at finite
string coupling.
\nxt
What is the analytical behavior of scattering amplitudes in 
non-zero RR background? Can one derive the identification
of RR fields directly from a discretization of super-Riemann 
surfaces?
\nxt
According to \cite{ghva}, the old-fashioned $c=1$ string at
the self-dual radius is related to the topological string 
on the conifold. It would be interesting to see whether our 
model with unequal chemical potentials corresponding to 
non-zero RR background could have a similar interpretation 
in the context of topological strings.

Needless to say, we hope to come back to some of these problems
in the future.

\begin{acknowledgments} 
It is a pleasure to thank
Tom Banks,
Oliver DeWolfe, 
Mike Douglas,
Jaume Gomis, 
Chris Herzog, 
Clifford Johnson, 
Don Marolf,
Joe Polchinski,
Radu Roiban,
Mark Spradlin,
and Anastasia Volovich
for helpful discussions and comments. 
This research was supported in part by the National Science Foundation under
Grant No.\ PHY99-07949.
\end{acknowledgments}

\begin{appendix}
\vskip 2em
\noindent 
{\bf\Large Appendix}

\section{$V(\tilde C)=V(C)$}
\label{app:S}

This discussion is {\it mutatis mutandis} almost identical to the one given
in section \ref{sec:defs} and \ref{sec:pot}. When the chemical potential 
$\mu$ is above the top of the matrix model potential, the perturbative 
energy levels are left/right-moving moving waves instead of localized on 
one side of the potential. The magnetic gauge symmetry associated with 
$\tilde C$ shifts the phase of left movers by $\ee^{\ii \tilde C}$ and 
the phase of right movers by $\ee^{-\ii\tilde C}$. We can implement
$\tilde C$ in the finite-$N$ matrix model by imposing periodic boundary
conditions on the wavefunctions with a phase shift of $\ee^{\ii \tilde C}$.
This means that we study the Schr\"odinger equation \eqref{parabolic}
and impose
\begin{equation}
\begin{split}
\ee^{\ii \tilde C/2} \psi(a,x_0) &= \ee^{-\ii \tilde C/2} \psi(a,-x_0) \\
\ee^{\ii \tilde C/2} \psi'(a,x_0)&= \ee^{-\ii \tilde C/2} \psi'(a,-x_0) \,,
\end{split}
\eqlabel{periodic}
\end{equation}
where $x_0 =\sqrt{2/\hbar}\to \infty$ in the double-scaling limit.
An alternative is to study the equation
\begin{equation}
\Bigl[\Bigl(\del_x-\ii \frac{\tilde C}{2 x_0}\Bigr)^2 + \frac{x^2}{4} - 
a\Bigr]\psi(a,x) = 0 \,.
\end{equation}
with strictly periodic boundary conditions on $\psi$. This description
is of course gauge equivalent to \eqref{periodic}, but makes it more
manifest that left/right movers pick up a phase shift when traversing
the potential from $-x_0$ to $x_0$. This implementation of $\tilde C$ 
is appropriate if we construct the matrix model using unitary matrices
with periodic potential.

Upon writing the general solution of \eqref{parabolic} in terms of
parabolic cylinder functions $W(a,x)$ and $W(a,-x)$, eq.\ 
\eqref{periodic} becomes the eigenvalue equation
\begin{equation}
\left|
\begin{matrix}
\ee^{\ii \tilde C/2} W(a,x) - \ee^{-\ii \tilde C/2} W(a,-x) &
\ee^{\ii \tilde C/2} W(a,-x) - \ee^{-\ii \tilde C/2} W(a,x) \\
\ee^{\ii \tilde C/2} \del_x W(a,x) + \ee^{-\ii \tilde C/2} \del_x 
\bigl(W(a,-x)\bigr) &
\ee^{\ii \tilde C/2} \del_x\bigl( W(a,-x)\bigr)+\ee^{-\ii \tilde C/2} 
\del_x W(a,x) 
\end{matrix}
\right| = 0 \,.
\eqlabel{determinant}
\end{equation}
Using the asymptotics for $x\gg |a|$,
\begin{equation}
W(a,x) \sim\sqrt{\frac{2k}x} \cos\varphi
\qquad
W(a,-x) \sim \sqrt{\frac{2}{k x}}\sin\varphi
\end{equation}
with $\varphi =\frac 14 x^2 - a\ln x + \frac\pi 4 + \frac 12\Phi_2$
and $k=\sqrt{1+\ee^{2\pi a}} - \ee^{\pi a}$, we obtain from 
\eqref{determinant}
\begin{equation}
\cos\tilde C = \sqrt{1+\ee^{2\pi a}} \sin 2\varphi_0 \,,
\eqlabel{eveq5}
\end{equation}
where $\varphi_0=\varphi$ for $x=x_0=\sqrt{2/\hbar}$. Up to a shift of 
$\varphi_0$ by $\pi/4$, this is exactly the ``S-dual'' of \eqref{eveq4} 
and shows that the non-perturbative potential one would compute for 
the magnetic variable $\tilde C$ is identical to the one we found 
for the electric variable $C$. Which potential is sensible to compute 
depends on the sign of $\mu$, \ie, whether we interpret $Q$ a electric 
or magnetic flux from the point of view of perturbative string theory.

\section{$V(C)$ at finite temperature} 
\label{app:V}

In this appendix, we compute the Fourier transform of $v'(a)$ 
in \eqref{vofa} in order to obtain a closed form expression for the
non-perturbative piece of the 0B partition function at finite
temperature and for general $C$. The quantity of interest is
\begin{equation}
\Delta'_{\rm np}(\mu) = \int_{-\infty}^\infty da\;
\frac1{2\pi} v'(a) f'(a,\mu) \,,
\end{equation}
where
\begin{equation}
f'(a,\mu) = \frac{\pi R}{2}\frac{1}{\cosh^2 \pi R(\mu-a)} \\
=\frac 1{2\pi} \int_{-\infty}^\infty dt\; \ee^{\ii(\mu-a)t}
\frac{t/ 2R}{\sinh t/2R} \,,
\end{equation}
and
\begin{equation}
\frac 1{2\pi} v'(a) = \frac{\ee^{-2\pi a}}{1+\ee^{-2\pi a}}
\,\frac{\cos C}{\sqrt{\sin^2C+\ee^{-2\pi a}}} \,.
\end{equation}
The integrand in
\begin{equation}
\int_{-\infty}^\infty da\; \ee^{\ii at} 
\frac{\ee^{-2\pi a}}{1+\ee^{-2\pi a}}\,
\frac{\cos C}{\sqrt{\sin^2C+\ee^{-2\pi a}}}
\eqlabel{fourier}
\end{equation}
has poles at $\ee^{-2\pi a}=-1$ and branch points at $\ee^{-2\pi a}=
-\sin^2 C$. It is convenient to place the cuts for $n\ge 0$ from 
$a=\bigl(n+\frac12\bigr)\ii-\frac 1{2\pi}\ln\sin^2C$ to 
$a=\bigl(n+\frac12\bigr)\ii +\infty$. We can then evaluate
the integral \eqref{fourier} for $t>0$ by closing the contour 
in the upper half plane, picking up the poles on the imaginary
axis and integrating back and forth along the cuts. Paying
attention to signs, we find the contribution from the
poles to be
\begin{equation}
\begin{split}
\sum_{n=0}^\infty 2\pi\ii \frac{-1}{2\pi} \ee^{-(n+\frac12)t}
\frac{\cos C}{\sqrt{\sin^2 C -1}}&=
\frac{\cos C}{|\cos C|} \sum_{n=0}^\infty (-1)^n \ee^{-(n+\frac12)t} \\
&=\frac{\cos C}{|\cos C|}\;\frac {1/2}{\cosh t/2} \,.
\end{split}
\end{equation}
To evaluate the contribution from the cuts, we substitute
$a=\bigl(n+\frac12\bigr)\ii -\frac 1{2\pi}\ln\sin^2C-\frac{1}{2\pi}\ln z$,
with $z\in[0,1]$, and are reduced to
\def\iit{\frac{\ii t}{2\pi}}
\begin{multline}
\sum_{n=0}^\infty  
\int_0^1 \frac {dz}z\;\frac{-2}{2\pi} (-1)^n
\ee^{-(n+\frac12)t} \bigl(\sin^2C\bigr)^{-\iit}
z^{-\iit} \frac{-z \sin^2C}{1-z\sin^2C} \frac{\cos C}{|\sin C|\sqrt{1-z}} \\
= \frac1\pi \; \frac{t/2}{\cosh t/2}\;|\sin C| \cos C \bigl(\sin^2 C
\bigr)^{-\iit} \int_0^1 dz\; z^{-\iit} (1-z)^{-1/2} (1-z\sin^2C)^{-1} \,.
\end{multline}
We recognize the remaining integral as the particular hypergeometric 
function
\begin{multline}
\frac{\Gamma\bigl(1-\iit\bigr)\Gamma\bigl(\frac12\bigr)}
{\Gamma\bigl(\frac 32-\iit\bigr)}
F\bigl(1,1-\iit;\frac32-\iit;\sin^2 C\bigr) \\ =
\Bigl(\frac12-\iit\Bigr)\;\frac{(\sin^2C)^{\iit}}{|\cos C|\,|\sin C|}\;
B\Bigl(1-\iit,\frac12\Bigr)\; B_{\sin^2 C}\Bigl(\frac12-\iit,\frac12\Bigr)
\end{multline}
which defines the incomplete Beta function according to the usual 
sources. Pulling things together, we can finally write
\begin{equation}
\begin{split}
\Delta'_{\rm np} (\mu)&= 
\frac 1{2\pi} \Re \int_0^\infty dt\; \ee^{-\ii\mu t} \,\frac{t/2R}
{\sinh t/2R}\,\frac{1/2}{\cosh t/2} \, \frac{\cos C}{|\cos C|}\,[ 1+ g(t,C)]\,,
\end{split}
\eqlabel{obpartfull}
\end{equation}
where
\begin{equation}
g(t,C) =\frac 1\pi \Bigl(\frac 12-\iit\Bigr)\;
B\Bigl(1-\iit,\frac12\Bigr)\; B_{\sin^2 C}\Bigl(\frac12-\iit,\frac12\Bigr) \,.
\end{equation}
The structure of the non-perturbative instanton contributions that follows
from these calculations is quite interesting. As we have seen in the main 
text \eqref{instantons}, the first instanton term for $C=0$ is of order 
$\ee^{-\pi\mu}$. On the other hand, when $C$ is non-zero, it can be seen
from \eqref{fourier} that non-perturbative effects set in at order
$\ee^{-2\pi\mu}$ for $\mu\to\infty$. (But diverge as $1/\sin C$ for $C\to 0$.)
This behavior is easy to understand in the matrix model. When $C=0$, the 
perturbative energy levels on the left and right match, and the simple 
tunneling amplitude is of order $\ee^{-\pi\mu}$. When $C\neq 0$, the 
perturbative levels do not match, and one has to tunnel forth and back 
to find a non-zero contribution.

\section{$\hat c=1$ at $R=1$}
\label{app:R}

In the bosonic $c=1$ matrix model, many formulas simplify at the self-dual
radius $R=1$. The supersymmetric model is not mapped onto itself under
$R\to 1/R$, as we have seen in the main text. But we can again evaluate 
some formulas more explicitly when we set $R=1$. For example, the 
perturbative part of the finite-temperature partition function is 
essentially the bosonic string result.
\begin{equation}
\begin{split}
\Delta''_{\rm pert}(\mu) &= \frac 1{2\pi} \int_0^\infty dt\;
\biggl(\frac {t/2}{\sinh t/2}\biggr)^2 \sin\mu t \\
&= \frac1{2\pi} \Re\,\del_\mu^2 \bigl(\mu\,\digamma(1\!-\!\ii\mu)\bigr)\,,
\end{split}
\end{equation}
leading to the well-known asymptotic expansion
\begin{equation}
2\pi \Delta''_{\rm pert}(\mu) \sim \frac 1\mu +
\sum_{n=1}^\infty \frac{(2n\!-\!1)|B_{2n}|}{\mu^{2n+1}} \,.
\end{equation}
In the present case, we of course have to add together two such
terms with $\mu=\mu_\pm$.

We can also compute the non-perturbative part of the partition function
exactly at $R=1$. Consider (again for one of $\mu_\pm$)
\begin{equation}
\begin{split}
\Delta_{\rm np}(\mu) &= \int_{-\infty}^\infty da\; \frac 1{2\pi} v'(a) f(a,\mu) \\
&=\int_{-\infty}^\infty da\; 
\frac1{1+\ee^{2\pi(\mu-a)}}\,
\frac{\ee^{-2\pi a}}{1+\ee^{-2\pi a}}\,
\frac{\cos C}{\sqrt{\sin^2C + \ee^{-2\pi a}}} \,, \\
\intertext{which setting $y=\ee^{2\pi\mu}$ and $z=\ee^{-2\pi a}$ can 
be reduced to an elementary integral}
&=\frac 1{2\pi}\int_0^\infty dz\;
\frac 1{1+y z} \,
\frac 1{1+z} \,
\frac{\cos C}{\sqrt{\sin^2 C+z}}\\
&= \frac{1}{2\pi} \frac{\cos C}{|\cos C|} 
\frac 1{y-1}
\left[ 2 \arcsin|\sin C|\! -\! \pi
- \sqrt{y} \,|\cos C|\,\frac{2\arcsin\bigl(|\sin C| \sqrt{y}\bigr)\!-\!\pi}
{\sqrt{1-y\sin^2 C}}\right] \,,
\end{split}
\end{equation}
where an appropriate branch of the $\arcsin$ is understood once $y\sin^2C >1$.

\end{appendix} 

\providecommand{\href}[2]{#2}\begingroup\raggedright


\begin{thebibliography}{1} 

\bibitem{gkp}
S.~B.~Giddings, S.~Kachru and J.~Polchinski,
``Hierarchies from fluxes in string compactifications,''
Phys.\ Rev.\ D {\bf 66}, 106006 (2002)
[arXiv:hep-th/0105097].

\bibitem{kklt}
S.~Kachru, R.~Kallosh, A.~Linde and S.~P.~Trivedi,
``De Sitter vacua in string theory,''
Phys.\ Rev.\ D {\bf 68}, 046005 (2003)
[arXiv:hep-th/0301240].

\bibitem{witten}
E.~Witten,
``Non-Perturbative Superpotentials In String Theory,''
Nucl.\ Phys.\ B {\bf 474}, 343 (1996)
[arXiv:hep-th/9604030].

\bibitem{mcve}
J.~McGreevy and H.~Verlinde,
``Strings from tachyons: The c = 1 matrix reloated,''
arXiv:hep-th/0304224.

\bibitem{tato}
T.~Takayanagi and N.~Toumbas,
``A matrix model dual of type 0B string theory in two dimensions,''
JHEP {\bf 0307}, 064 (2003)
[arXiv:hep-th/0307083].

\bibitem{dkkmms}
M.~R.~Douglas, I.~R.~Klebanov, D.~Kutasov, J.~Maldacena, E.~Martinec and N.~Seiberg,
``A new hat for the c = 1 matrix model,''
arXiv:hep-th/0307195.

\bibitem{drsvw}
O.~DeWolfe, R.~Roiban, M.~Spradlin, A.~Volovich and J.~Walcher,
``On the S-matrix of type 0 string theory,''
arXiv:hep-th/0309148.

\bibitem{kapustin}
A.~Kapustin,
``Noncritical superstrings in a Ramond-Ramond background,''
arXiv:hep-th/0308119.

\bibitem{gross}
D.~J.~Gross,
``The C = 1 Matrix Models,''
\href{http://www.slac.stanford.edu/spires/find/hep/www?irn=2698811}{SPIRES entry}

\bibitem{klebanov}
I.~R.~Klebanov,
``String theory in two-dimensions,''
arXiv:hep-th/9108019.

\bibitem{gimo}
P.~Ginsparg and G.~W.~Moore,
``Lectures On 2-D Gravity And 2-D String Theory,''
arXiv:hep-th/9304011.

\bibitem{polchinski}
J.~Polchinski,
``What is string theory?,''
arXiv:hep-th/9411028.

\bibitem{gv1}
A.~Sen,
``Open and closed strings from unstable D-branes,''
arXiv:hep-th/0305011.

\bibitem{gv2}
E.~J.~Martinec,
``The annular report on non-critical string theory,''
arXiv:hep-th/0305148.

\bibitem{kms}
I.~R.~Klebanov, J.~Maldacena and N.~Seiberg,
``D-brane decay in two-dimensional string theory,''
JHEP {\bf 0307}, 045 (2003)
[arXiv:hep-th/0305159].

\bibitem{gv3}
N.~R.~Constable and F.~Larsen,
``The rolling tachyon as a matrix model,''
JHEP {\bf 0306}, 017 (2003)
[arXiv:hep-th/0305177].

\bibitem{gv4}
J.~McGreevy, J.~Teschner and H.~Verlinde,
``Classical and quantum D-branes in 2D string theory,''
arXiv:hep-th/0305194.

\bibitem{gv5}
V.~Schomerus,
``Rolling tachyons from Liouville theory,''
arXiv:hep-th/0306026.

\bibitem{gv6}
S.~Y.~Alexandrov, V.~A.~Kazakov and D.~Kutasov,
``Non-perturbative effects in matrix models and D-branes,''
JHEP {\bf 0309}, 057 (2003)
[arXiv:hep-th/0306177].

\bibitem{gv7}
D.~Gaiotto, N.~Itzhaki and L.~Rastelli,
``On the BCFT description of holes in the c = 1 matrix model,''
arXiv:hep-th/0307221.

\bibitem{gv8}
M.~Gutperle and P.~Kraus,
``D-brane dynamics in the c = 1 matrix model,''
arXiv:hep-th/0308047.

\bibitem{gv9}
A.~Sen,
``Open-closed duality: Lessons from matrix model,''
arXiv:hep-th/0308068.

\bibitem{mmv}
J.~McGreevy, S.~Murthy and H.~Verlinde,
``Two-dimensional superstrings and the supersymmetric matrix model,''
arXiv:hep-th/0308105.

\bibitem{gv10}
A.~Giveon, A.~Konechny, A.~Pakman and A.~Sever,
``Type 0 strings in a 2-d black hole,''
arXiv:hep-th/0309056.

\bibitem{gv11}
J.~L.~Karczmarek and A.~Strominger,
``Matrix cosmology,''
arXiv:hep-th/0309138.

\bibitem{gv12}
I.~R.~Klebanov, J.~Maldacena and N.~Seiberg,
``Unitary and complex matrix models as 1-d type 0 strings,''
arXiv:hep-th/0309168.

\bibitem{gv13}
S.~Ribault and V.~Schomerus,
``Branes in the 2-D black hole,''
arXiv:hep-th/0310024.

\bibitem{gv14}
S.~Dasgupta and T.~Dasgupta,
``Renormalization group approach to c = 1 matrix model on a circle and D-brane decay,''
arXiv:hep-th/0310106.

\bibitem{gv15}
S.~Alexandrov,
``(m,n) ZZ branes and the c = 1 matrix model,''
arXiv:hep-th/0310135.

\bibitem{gv16}
J.~Gomis and A.~Kapustin,
``Two-dimensional unoriented strings and matrix models,''
arXiv:hep-th/0310195.

\bibitem{brka}
E.~Brezin and V.~A.~Kazakov,
``Exactly Solvable Field Theories Of Closed Strings,''
Phys.\ Lett.\ B {\bf 236}, 144 (1990).

\bibitem{dosh}
M.~R.~Douglas and S.~H.~Shenker,
``Strings In Less Than One-Dimension,''
Nucl.\ Phys.\ B {\bf 335}, 635 (1990).

\bibitem{grmi}
D.~J.~Gross and A.~A.~Migdal,
``Nonperturbative Two-Dimensional Quantum Gravity,''
Phys.\ Rev.\ Lett.\  {\bf 64}, 127 (1990).

\bibitem{bega}
O.~Bergman and M.~R.~Gaberdiel,
``A non-supersymmetric open-string theory and S-duality,''
Nucl.\ Phys.\ B {\bf 499}, 183 (1997)
[arXiv:hep-th/9701137].

\bibitem{klts}
I.~R.~Klebanov and A.~A.~Tseytlin,
``D-branes and dual gauge theories in type 0 strings,''
Nucl.\ Phys.\ B {\bf 546}, 155 (1999)
[arXiv:hep-th/9811035].

\bibitem{klts2}
I.~R.~Klebanov and A.~A.~Tseytlin,
``A non-supersymmetric large N CFT from type 0 string theory,''
JHEP {\bf 9903}, 015 (1999)
[arXiv:hep-th/9901101].

\bibitem{meor}
P.~Meessen and T.~Ortin,
``Type 0 T-duality and the tachyon coupling,''
Phys.\ Rev.\ D {\bf 64}, 126005 (2001)
[arXiv:hep-th/0103244].

\bibitem{thompson}
D.~M.~Thompson,
``Descent relations in type 0A / 0B,''
Phys.\ Rev.\ D {\bf 65}, 106005 (2002)
[arXiv:hep-th/0105314].

\bibitem{grkl}
D.~J.~Gross and I.~R.~Klebanov,
``One-Dimensional String Theory On A Circle,''
Nucl.\ Phys.\ B {\bf 344}, 475 (1990).

\bibitem{moore}
G.~W.~Moore,
``Double scaled field theory at c = 1,''
Nucl.\ Phys.\ B {\bf 368} (1992) 557.

\bibitem{mpr}
G.~W.~Moore, M.~R.~Plesser and S.~Ramgoolam,
``Exact S matrix for 2-D string theory,''
Nucl.\ Phys.\ B {\bf 377}, 143 (1992)
[arXiv:hep-th/9111035].

\bibitem{tachyon}
A.~Sen,
``Non-BPS states and branes in string theory,''
arXiv:hep-th/9904207.

\bibitem{jeyo}
A.~Jevicki and T.~Yoneya,
``A Deformed matrix model and the black hole background in
two-dimensional string theory,''
Nucl.\ Phys.\ B {\bf 411}, 64 (1994)
[arXiv:hep-th/9305109].

\bibitem{dkr}
K.~Demeterfi, I.~R.~Klebanov and J.~P.~Rodrigues,
``The Exact S matrix of the deformed c = 1 matrix model,''
Phys.\ Rev.\ Lett.\  {\bf 71}, 3409 (1993)
[arXiv:hep-th/9308036].

\bibitem{ZZ}
A.~B.~Zamolodchikov and A.~B.~Zamolodchikov,
``Liouville field theory on a pseudosphere,''
arXiv:hep-th/0101152.

\bibitem{abst}
M.~Abramowitz and I.~A.~Stegun, ``Handbook of Mathematical Functions,''
9th Printing, Dover Publications, Inc., New York 1970.

\bibitem{aspl}
P.~S.~Aspinwall and M.~R.~Plesser,
``T-duality can fail,''
JHEP {\bf 9908}, 001 (1999)
[arXiv:hep-th/9905036].

\bibitem{ghva}
D.~Ghoshal and C.~Vafa,
``C = 1 string as the topological theory of the conifold,''
Nucl.\ Phys.\ B {\bf 453}, 121 (1995)
[arXiv:hep-th/9506122].


\end{thebibliography}
\end{document}